\documentclass[paper]{JHEP}
\usepackage{epsfig}
\def\be{\begin{equation}}
\def\ee{\end{equation}}
\def\bea{\begin{eqnarray}}
\def\eea{\end{eqnarray}}
\def\rarr{\rightarrow}

\def\kf{{\bf k}}
\def\qf{{\bf q}}
\def\lf{{\bf l}}
\def\C{{\rm\kern.24em
    \vrule width.02em height1.4ex depth-.05ex
    \kern-.26em C}}
\def\N{{\rm I\kern-.18em N}}
\def\O{{\rm\kern.24em
    \vrule width.02em height1.4ex depth-.05ex
    \kern-.26em O}}
\def\P{{\rm I\kern-.25em P}}
\def\R{{\rm I\kern-.21em R}}
\def\Z{{\rm\kern.26em
    \vrule width.02em height0.5ex depth 0ex
    \kern.04em
    \vrule width.02em height1.47ex depth-1ex
    \kern-.34em Z}}
\def\nn{\nonumber}
\def\fr{\frac}

\newdimen\picraise
\newcommand\picbox[1]
{
  \setbox0=\hbox{\input{#1}}
  \picraise=-0.5\ht0 
  \advance\picraise by 0.5\dp0
  \advance\picraise by 3pt     % correct for height of `=' above baseline
  \hbox{\raise\picraise \box0}
}
\newdimen\picraiset
\newcommand\picding[1]
{
  \setbox0=\hbox{\input{#1}}
  \picraiset=-0.5\ht0 
  \advance\picraiset by 0.5\dp0
  \advance\picraiset by -3pt  % correct for height of pictures in inteq. 
  \hbox{\raise\picraiset \box0}
}
\newdimen\picraisehallo
\newcommand\pichallo[2]
{
  \setbox0=\hbox{\input{#1}}
  \picraisehallo=-0.5\ht0 
  \advance\picraisehallo by 0.5\dp0
  \advance\picraisehallo by -#2pt  % correct for height of pictures in inteq. 
  \hbox{\raise\picraisehallo \box0}
}
\title{Reggeization in High Energy QCD
\thanks{Work supported in part by the EU Fourth Framework Programme
`Training and Mobility of Researchers', Network `Quantum Chromodynamics
and the Deep Structure of Elementary Particles', 
contract FMRX-CT98-0194 (DG 12 - MIHT).}
}

\author{Carlo Ewerz\\
             Institut f\"ur Theoretische Physik, Universit\"at Heidelberg\\
             Philosophenweg 16, D-69120 Heidelberg, 
             Germany\\
             E-mail: \email{carlo@thphys.uni-heidelberg.de}}

\abstract{
We study QCD in the regime of high parton density
arising in hadronic collisions at large center--of--mass energy. 
The $n$-gluon amplitudes of 
the generalized leading logarithmic approximation 
are investigated. We find 
identities relating amplitudes with different 
numbers of gluons in the $t$-channel. 
These identities constrain  
the reggeization of the gluon in high energy QCD. 
The tensors describing the reggeization of a gluon 
in color space are identified. 
}


\preprint{HD-THEP-01-12}

\begin{document} 

\section{Introduction}
\label{intro}
A longstanding problem in the physics of strong interactions 
is to understand the Regge limit of hadronic scattering amplitudes. 
Especially challenging is the description of this kinematical region 
of large center--of--mass energy $\sqrt{s}$ and small 
momentum transfer $\sqrt{t}$ in terms of quark and gluon degrees 
of freedom. Regge theory provides a successful phenomenology 
but it has not yet been derived from QCD. The main problem 
is the fact that hadronic scattering in the Regge limit is 
in general characterized  
by large parton densities and small momentum scales. A full 
understanding will eventually require non--perturbative 
methods. 
Scattering processes involving small color dipoles, however, 
can be described using perturbative methods. Examples are 
heavy onium scattering or collisions of highly virtual photons. 
These processes involve a large momentum scale (the heavy quark 
mass or the photon virtuality, respectively) and can be treated 
perturbatively even at high energy. We therefore hope that by 
studying these processes some essential features of the dynamics 
of high energy QCD can be discovered using perturbative methods. 

The starting point of the perturbative approach to high energy QCD 
is the BFKL Pomeron \cite{FKL,BL}. It resums terms of the 
form $(\alpha_s \log s)^n$ in which large logarithms of the energy 
can compensate the smallness of the strong coupling constant $\alpha_s$. 
The corresponding approximation scheme is known as the leading 
logarithmic approximation (LLA). The BFKL Pomeron leads to 
a power--like growth of total cross sections with energy, 
$\sigma \sim s^\lambda$, 
where $\lambda = 4 \alpha_s N_c \ln 2 /\pi$. 
On the level of the total cross section 
this perturbative Pomeron can be interpreted as an exchange of a 
ladder of gluons in the $t$-channel. In other words: the BFKL 
equation describes the exchange of a bound state of two gluons 
in the $t$-channel. 

A very important outcome of the perturbative (BFKL) analysis 
of high energy QCD is the so--called reggeization of the gluon 
\cite{Reggeization}. Technically speaking it manifests itself 
in the fact that the BFKL equation in the color octet channel 
exhibits a pole solution, corresponding to a single (reggeized) particle 
with gluon quantum numbers propagating in the $t$-channel. 
Looking at this fact from another angle, 
we see that a gluon propagating in the $t$-channel 
turns out to be a composite object, namely a bound state 
of two gluons. This apparently confusing situation 
becomes quite natural when we realize that 
elementary gluons are just not the most suitable 
degrees of freedom. Instead, the natural degrees of freedom 
in the high energy limit are collective excitations of the 
non--abelian gauge field, i.\,e.\ the reggeized gluons or reggeons. 
The elementary gluon and the bound state of two 
gluons in a color octet state mentioned above can then 
be interpreted as the first two terms in an expansion 
of the reggeized gluon in the number of its constituents, 
similar to the Fock states of a usual bound state. 
It is the purpose of the present paper to study this 
expansion in more detail. We will in particular 
investigate the behavior of color quantum 
numbers in the process of reggeization.  

The most suitable framework for studying these 
aspects of reggeization is the generalized leading logarithmic 
approximation (GLLA) 
\cite{Bartelsnuclphys,Bartelskernels,Bartelsinteq} 
which is a natural extension of the 
BFKL formalism to larger numbers of gluons in 
the $t$-channel. It has been widely studied in the 
context of unitarity corrections. These unitarity 
corrections and the GLLA have mostly been analysed for the 
process of virtual photon collisions, but the results 
are expected to apply to a variety of processes. 
In collisions at very 
large energies the parton densities become large. 
Consequently, parton recombination effects become important 
and the rapid growth of cross sections with 
energy will be slowed down. 
The GLLA has been devised to incorporate these 
recombination effects. It is most conveniently formulated 
in terms of multi--gluon amplitudes describing the 
production of a number of gluons in the $t$-channel. 
During the evolution in the $t$-channel 
the number of gluons is allowed 
to fluctuate and thus transitions between different 
$n$-gluon states are possible. The two--gluon 
amplitude in this framework is just the usual 
BFKL amplitude. 

The amplitudes for the production of up to four 
gluons in the $t$-channel have been analysed in 
\cite{BPLB,BW,Hans,Braun,Vacca,CE}. They turn out to 
have a very interesting field theory structure 
which manifests itself in the following way. 
In these amplitudes there is a two--gluon 
state (the BFKL Pomeron) 
as well as a state of four interacting gluons in the 
$t$-channel. In addition, there is a transition 
vertex $V_{2 \rarr 4}$ coupling these states to 
each other. The latter is a number--changing 
vertex and thus turns the quantum mechanical 
system of $n$-gluon states into a quantum field 
theory of interacting $n$-gluon states. Interestingly, 
only states of even numbers of gluons occur. The amplitudes 
are functions of the two--dimensional transverse momenta 
of the gluons. They also depend on rapidity which 
can be understood as a time--like variable. After a 
Fourier transformation to two--dimensional 
impact parameter space the amplitudes can be 
shown to be invariant under conformal transformations 
of the gluon coordinates \cite{Lip86,BLW}. 
In \cite{BE} also the amplitudes with up to six 
gluons have been studied. It was shown 
that the field theory structure persists also to larger 
numbers of gluons and the transition vertex from 
a two--gluon to a six--gluon state was calculated. 
Also this vertex is conformally invariant in impact 
parameter space \cite{inprep}. 
In summary, the $n$-gluon amplitudes have the 
structure of a conformally invariant field theory in 
two--dimensional impact parameter space with rapidity 
as an additional time--like parameter. 
In its present form this field theory is formulated 
in terms of $n$-gluon states and vertices. It would 
obviously be desirable to extract from these the 
necessary information to identify the corresponding 
two--dimensional conformal field theory and 
to apply the powerful methods available for the latter. 
At present this important step is an open problem. 
A good understanding of how the field theory structure 
emerges should be helpful to make 
progress in this direction. 
We expect reggeization to play a significant r\^ole 
in this context because it has been found to be 
an indispensable condition for the emergence of the 
field theory structure \cite{BE}. 
Especially the details of reggeization in color space 
are expected to contain important information.

The $n$-gluon amplitudes have been investigated for up to 
six $t$-channel gluons. Moreover, a certain 
part of the amplitudes can be computed even for 
arbitrary $n$, the result being a superposition of 
BFKL amplitudes. These results are explained in 
some detail in section \ref{review} where we also 
collect some formulae and notation needed in this paper. 
Equipped with these (partial) solutions 
we are in a position to study two novel aspects of their structure 
related to reggeization. 
Both concern the intricate interplay of color and momentum 
structure present in the amplitudes. 
The first aspect is the subject of section \ref{Ward}. It is of 
a more global nature and relates $n$-gluon amplitudes of 
different $n$. The corresponding identities are obtained when 
one of the gluon momenta vanishes. They place strong 
constraints on the color structure of the complete amplitude. 
In section \ref{tensors} we encounter 
more local properties, namely the color tensors accompanying 
the reggeization of a single gluon. 
Our aim is to extract from the known amplitudes 
as much information as possible. 
We then formulate conjectures on how the 
observations made here can be generalized to a larger number 
of $t$-channel gluons or to the unknown pieces of the 
amplitudes, respectively. We also discuss the relevance 
and the potential uses of the conjectures for the 
further investigation of the effective field theory of unitarity 
corrections. 

\boldmath
\section{Multi-gluon amplitudes and reggeization}
\unboldmath
\label{review}

In this section we collect results which we will need 
in the subsequent sections. For a more extensive 
review and the detailed derivation of these results the 
reader is referred to \cite{BE} and references therein. 

\subsection{BFKL equation and reggeization}

The BFKL equation for the partial wave 
amplitude $\phi_\omega$ reads 
\be
  \omega \phi_\omega(\kf_1,\kf_2) = 
 \phi^0(\kf_1,\kf_2) 
 + \int \frac{d^2\lf}{(2\pi)^3} \,
 \frac{1}{\lf^2 (\qf-\lf)^2} 
 K_{\mbox{\scriptsize BFKL}}(\lf,\qf-\lf;\kf_1,\kf_2) 
\,\phi_\omega(\lf,\qf-\lf) 
\,.
\label{BFKLeq}
\ee
Here $\kf_1$ and $\kf_2$ are the transverse momenta 
of the two gluons in the $t$-channel, and $q=\kf_1+\kf_2$ is 
the total transverse momentum transfer in the $t$-channel. 
$\omega$ denotes the complex angular momentum 
conjugate to rapidity. $\phi^0$ is an inhomogeneous 
term depending on the process under consideration 
which couples the two gluons to external particles.  
$K_{\mbox{\scriptsize BFKL}}$ is the Lipatov kernel, 
\bea
\label{Lipatovkernel}
K_{\mbox{\scriptsize BFKL}}(\lf,\qf-\lf;\kf,\qf-\kf) &=& 
 -  N_c g^2 \left[ \qf^2 - \frac{\kf^2(\qf-\lf)^2}{(\kf-\lf)^2} 
 - \frac{(\qf-\kf)^2 \lf^2}{(\kf-\lf)^2} \right] \nonumber \\
 & & +(2\pi)^3 \kf^2 (\qf-\kf)^2 
 \left[ \,\beta(\kf) + \beta(\qf - \kf) \right] 
 \delta^{(2)}(\kf-\lf) 
\eea
with the gluon trajectory function 
$\alpha(\kf^2) = 1 + \beta(\kf^2)$ where 
\be
  \beta(\kf^2) = \frac{N_c}{2} g^2  \int \frac{d^2\lf}{(2 \pi)^3} 
          \frac{\kf^2}{\lf^2 (\lf -\kf)^2} 
\,.
\label{traject}
\ee

If the two gluons in the $t$-channel are in a color octet state 
the factor $-N_c$ in the kernel (\ref{Lipatovkernel}) has 
to be replaced by $-N_c/2$. Assuming that $\phi_0$ depends 
only on the sum $q=\kf_1+\kf_2$ of the momenta 
of the two $t$-channel gluons we find that the BFKL equation 
(\ref{BFKLeq}) has a pole solution 
\be
\phi^{\bf 8_A} (\kf_1+\kf_2) = 
\frac{\phi^{\bf 8_A}_0 (\kf_1+\kf_2)}{\omega -\beta(\kf_1+\kf_2)} 
\,.
\label{Pole}
\ee
As already discussed in the Introduction it signals the reggeization 
of the gluon, i.\,e.\ it indicates that the color octet exchange is 
a composite object of two gluons. 
In addition, the amplitude $\phi^{\bf 8_A}$ can 
be `squared' in the $t$-channel to give a result proportional 
to the trajectory function $\beta(\qf^2)$ (see (\ref{traject})) 
of the reggeized gluon. In the present paper, however, we will not 
make use of the corresponding $t$-channel unitarity relations and 
concentrate on the amplitudes describing the production 
of $2$ (or in general $n$) gluons in the $t$-channel. 

\subsection{Multigluon amplitudes}
\label{reviewmultigluon}
The multigluon amplitudes $D_n$ we will consider from 
now on apply to the process of virtual photon--photon 
scattering. They are cut in the $t$-channel and thus 
describe the production of $n$ gluons in the 
$t$-channel starting from two virtual photons. 
Similar to the BFKL equation they 
are partial wave amplitudes depending on the complex 
angular momentum $\omega$ (which we suppress in 
the notation). They are also cut in the $n-1$ energy variables 
formed from the first photon and the $i$ first gluons 
($1 \le i \le n$). The amplitudes depend on the $n$ 
transverse momenta $\kf_i$ of the gluons and 
on their color labels $a_i$. 
These amplitudes $D_n$ are obtained as solutions of 
a tower of coupled integral equations. The equation 
for $n=2$ is just the BFKL equation with the particular 
inhomogeneous term induced by the coupling of two 
gluons to the virtual photons. 
In \cite{BE} the integral equations have partly been 
solved. Before giving the solutions explicitly we first introduce 
suitable tensors and notation in color space. 

We consider the gauge group $SU(N_c)$ with 
generators $t^a$ ($a=1,\dots,N_c^2-1$). The algebra is 
\be
  [ t^a, t^b ] = i f_{abc} t^c  \,.  
\label{algebra}
\ee
and for the structure constants $f_{abc}$ 
\be
   f_{abc} = - f_{acb} = 
             - 2 i \, [ \mbox{tr} ( t^a t^b t^c ) 
                          - \mbox{tr} ( t^c t^b t^a ) ] \,.
\label{fabcdef}
\ee
Introducing birdtrack notation \cite{Cvitanovic} 
they become 
\be
  f_{abc} = 
     \begin{picture}(0,0)%
\includegraphics{f_abc.pstex}%
\end{picture}%
\setlength{\unitlength}{3947sp}%
\begingroup\makeatletter\ifx\SetFigFont\undefined
% extract first six characters in \fmtname
\def\x#1#2#3#4#5#6#7\relax{\def\x{#1#2#3#4#5#6}}%
\expandafter\x\fmtname xxxxxx\relax \def\y{splain}%
\ifx\x\y   % LaTeX or SliTeX?
\gdef\SetFigFont#1#2#3{%
  \ifnum #1<17\tiny\else \ifnum #1<20\small\else
  \ifnum #1<24\normalsize\else \ifnum #1<29\large\else
  \ifnum #1<34\Large\else \ifnum #1<41\LARGE\else
     \huge\fi\fi\fi\fi\fi\fi
  \csname #3\endcsname}%
\else
\gdef\SetFigFont#1#2#3{\begingroup
  \count@#1\relax \ifnum 25<\count@\count@25\fi
  \def\x{\endgroup\@setsize\SetFigFont{#2pt}}%
  \expandafter\x
    \csname \romannumeral\the\count@ pt\expandafter\endcsname
    \csname @\romannumeral\the\count@ pt\endcsname
  \csname #3\endcsname}%
\fi
\fi\endgroup
\begin{picture}(675,597)(1126,-298)
\put(1801,-286){\makebox(0,0)[lb]{\smash{\SetFigFont{10}{12.0}{rm}$c$}}}
\put(1351,164){\makebox(0,0)[lb]{\smash{\SetFigFont{10}{12.0}{rm}$a$}}}
\put(1126,-286){\makebox(0,0)[lb]{\smash{\SetFigFont{10}{12.0}{rm}$b$}}}
\end{picture}
 =  \,- \hspace{-.2cm} \begin{picture}(0,0)%
\includegraphics{f_acb.pstex}%
\end{picture}%
\setlength{\unitlength}{3947sp}%
\begingroup\makeatletter\ifx\SetFigFont\undefined
% extract first six characters in \fmtname
\def\x#1#2#3#4#5#6#7\relax{\def\x{#1#2#3#4#5#6}}%
\expandafter\x\fmtname xxxxxx\relax \def\y{splain}%
\ifx\x\y   % LaTeX or SliTeX?
\gdef\SetFigFont#1#2#3{%
  \ifnum #1<17\tiny\else \ifnum #1<20\small\else
  \ifnum #1<24\normalsize\else \ifnum #1<29\large\else
  \ifnum #1<34\Large\else \ifnum #1<41\LARGE\else
     \huge\fi\fi\fi\fi\fi\fi
  \csname #3\endcsname}%
\else
\gdef\SetFigFont#1#2#3{\begingroup
  \count@#1\relax \ifnum 25<\count@\count@25\fi
  \def\x{\endgroup\@setsize\SetFigFont{#2pt}}%
  \expandafter\x
    \csname \romannumeral\the\count@ pt\expandafter\endcsname
    \csname @\romannumeral\the\count@ pt\endcsname
  \csname #3\endcsname}%
\fi
\fi\endgroup
\begin{picture}(675,597)(1126,-298)
\put(1801,-286){\makebox(0,0)[lb]{\smash{\SetFigFont{10}{12.0}{rm}$c$}}}
\put(1351,164){\makebox(0,0)[lb]{\smash{\SetFigFont{10}{12.0}{rm}$a$}}}
\put(1126,-286){\makebox(0,0)[lb]{\smash{\SetFigFont{10}{12.0}{rm}$b$}}}
\end{picture}

     =  - 2 i \left[ \, \begin{picture}(0,0)%
\includegraphics{trabc.pstex}%
\end{picture}%
\setlength{\unitlength}{3947sp}%
\begingroup\makeatletter\ifx\SetFigFont\undefined
% extract first six characters in \fmtname
\def\x#1#2#3#4#5#6#7\relax{\def\x{#1#2#3#4#5#6}}%
\expandafter\x\fmtname xxxxxx\relax \def\y{splain}%
\ifx\x\y   % LaTeX or SliTeX?
\gdef\SetFigFont#1#2#3{%
  \ifnum #1<17\tiny\else \ifnum #1<20\small\else
  \ifnum #1<24\normalsize\else \ifnum #1<29\large\else
  \ifnum #1<34\Large\else \ifnum #1<41\LARGE\else
     \huge\fi\fi\fi\fi\fi\fi
  \csname #3\endcsname}%
\else
\gdef\SetFigFont#1#2#3{\begingroup
  \count@#1\relax \ifnum 25<\count@\count@25\fi
  \def\x{\endgroup\@setsize\SetFigFont{#2pt}}%
  \expandafter\x
    \csname \romannumeral\the\count@ pt\expandafter\endcsname
    \csname @\romannumeral\the\count@ pt\endcsname
  \csname #3\endcsname}%
\fi
\fi\endgroup
\begin{picture}(675,660)(676,-361)
\put(1351,-361){\makebox(0,0)[lb]{\smash{\SetFigFont{10}{12.0}{rm}$c$}}}
\put(901,164){\makebox(0,0)[lb]{\smash{\SetFigFont{10}{12.0}{rm}$a$}}}
\put(676,-361){\makebox(0,0)[lb]{\smash{\SetFigFont{10}{12.0}{rm}$b$}}}
\end{picture}
 - 
                  \,\begin{picture}(0,0)%
\includegraphics{tracb.pstex}%
\end{picture}%
\setlength{\unitlength}{3947sp}%
\begingroup\makeatletter\ifx\SetFigFont\undefined
% extract first six characters in \fmtname
\def\x#1#2#3#4#5#6#7\relax{\def\x{#1#2#3#4#5#6}}%
\expandafter\x\fmtname xxxxxx\relax \def\y{splain}%
\ifx\x\y   % LaTeX or SliTeX?
\gdef\SetFigFont#1#2#3{%
  \ifnum #1<17\tiny\else \ifnum #1<20\small\else
  \ifnum #1<24\normalsize\else \ifnum #1<29\large\else
  \ifnum #1<34\Large\else \ifnum #1<41\LARGE\else
     \huge\fi\fi\fi\fi\fi\fi
  \csname #3\endcsname}%
\else
\gdef\SetFigFont#1#2#3{\begingroup
  \count@#1\relax \ifnum 25<\count@\count@25\fi
  \def\x{\endgroup\@setsize\SetFigFont{#2pt}}%
  \expandafter\x
    \csname \romannumeral\the\count@ pt\expandafter\endcsname
    \csname @\romannumeral\the\count@ pt\endcsname
  \csname #3\endcsname}%
\fi
\fi\endgroup
\begin{picture}(675,660)(1126,-361)
\put(1801,-361){\makebox(0,0)[lb]{\smash{\SetFigFont{10}{12.0}{rm}$c$}}}
\put(1351,164){\makebox(0,0)[lb]{\smash{\SetFigFont{10}{12.0}{rm}$a$}}}
\put(1126,-361){\makebox(0,0)[lb]{\smash{\SetFigFont{10}{12.0}{rm}$b$}}}
\end{picture}
 \, \right] \,,
\ee
in which the orientated lines denote quark color representations whereas 
the unorientated lines correspond to gluon color lines. 
It will be useful to define the color tensors 
\bea
  d^{b_1 b_2 \dots b_n} &=& 
           \mbox{tr} (t^{b_1} t^{b_2} \dots t^{b_n} )  
         + \mbox{tr} (t^{b_n} \dots t^{b_2} t^{b_1} ) \label{dallgdef} \\
  f^{b_1 b_2 \dots b_n} &=& 
    \fr{1}{i} \,  [ \mbox{tr} (t^{b_1} t^{b_2}\dots t^{b_n}) 
         - \mbox{tr} (t^{b_n} \dots t^{b_2} t^{b_1} ) ] 
    \label{fallgdef} 
\,.
\eea
These tensors are obviously invariant under cyclic 
permutations of the color labels. 
To make contact with the usual structure constants (which 
we write with lower indices) we mention that 
$d^{abc} = \frac{1}{2} d_{abc}$ and $ f^{abc} = \frac{1}{2} f_{abc}$. 
Further we have $d^{ab} = \delta_{ab}$. 

It will also be useful to recall some facts about invariant tensors. 
The $n$ gluons of our amplitudes $D_n$ form 
an overall color singlet. They can therefore be 
expanded in color space into a linear combination of 
invariant $\mbox{su}(N_c)$ tensors. 
Invariant tensors in a simple Lie algebra can 
generally be constructed from traces of group generators, 
\be
  \Theta^{a_1\dots a_n} = \mbox{tr}(t^{a_1} \dots t^{a_n}) 
\,.
\ee
For such a trace the property of being an invariant tensor reads  
\be
  \sum_{i=1}^n f_{ca_ib} \,
  \Theta^{a_1\dots b \dots a_n} = 0
\,,
\label{invariancecond}
\ee
$t^b$ being inserted at the $i$th position in the tensor 
(i.\,e.\ in the trace). Equation (\ref{invariancecond}) is often 
used as the defining property of an invariant tensor. 
Our tensors defined in eqs.\ (\ref{dallgdef}) and (\ref{fallgdef}) 
are obtained as sum or difference 
of such traces and thus invariant tensors fulfilling 
the condition (\ref{invariancecond}). 

The integral equations describing the $n$-gluon amplitudes 
contain as a lowest order term the coupling $D_{(n;0)}$ of $n$ gluons 
to the two virtual photons via a quark loop, as shown in Fig.\ 
\ref{fig:nocrossing} for the case of four gluons. 
\FIGURE{
\begin{picture}(0,0)%
\epsfig{file=nocross.pstex}%
\end{picture}%
\setlength{\unitlength}{3947sp}%
\begingroup\makeatletter\ifx\SetFigFont\undefined%
\gdef\SetFigFont#1#2#3#4#5{%
  \reset@font\fontsize{#1}{#2pt}%
  \fontfamily{#3}\fontseries{#4}\fontshape{#5}%
  \selectfont}%
\fi\endgroup%
\begin{picture}(5102,1224)(600,-1798)
\end{picture}

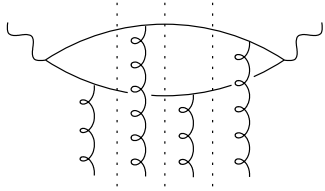
\caption{Cut amplitude contributing to the coupling of $n=4$ 
gluons to a quark loop}
\label{fig:nocrossing}}
The $n$ gluons are attached to the quark loop in all possible 
ways in order to preserve gauge invariance, but due to the 
cuts (indicated in the figure by dashed lines) the gluons 
are not allowed to cross in the $t$-channel. 
The term $D_{(n;0)}$ is thus obtained as the sum of 
$2^n$ diagrams since each of the gluons can be coupled 
either to the quark or to the antiquark. 
%It can be shown that 
%the two terms in which all gluons couple to the same 
%(quark or antiquark) line act as regularization terms. 
As discussed in detail in \cite{BE} the lowest order amplitude 
$D_{(n;0)}$ can be expressed in terms of $D_{(2;0)}$ only. 
The coupling of three gluons to the quark loop is for example 
given by
\be
  D_{(3;0)}^{a_1a_2a_3}(\kf_1,\kf_2,\kf_3) 
= \fr{1}{2} g f_{a_1a_2a_3} \,
[ D_{(2;0)}(12,3) - D_{(2;0)}(13,2) + D_{(2;0)}(1,23) ]  \,. 
\label{d30}
\ee
Here we have introduced a new notation for the momentum 
arguments: for brevity we replace the momentum $\kf_i$ by 
its index, and a string of indices stands for to the sum of the 
corresponding momenta such that for example 
\be
  D_{(2;0)}(12,3) = D_{(2;0)}(\kf_1+\kf_2,\kf_3) \,. 
\ee
The fact that $D_{(3;0)}$ the eight diagrams reduce to three terms 
only is due to the fact that two of the diagrams play a 
special r\^ole: the diagrams in which all gluons are coupled 
to the same quark (or antiquark) line act as regularization 
terms only and do not lead to an extra term in (\ref{d30}). 

We now turn to the full $n$-gluon amplitudes obtained as 
solutions of the integral equations. As already mentioned the 
two gluon amplitude $D_2$ is just the well--known BFKL 
amplitude with the photon impact factor as the initial condition 
of the $t$-channel evolution. The color structure of $D_2$ is 
trivial, 
\be
  D_2^{a_1a_2} (\kf_1,\kf_2) = \delta_{a_1a_2} D_2 (\kf_1,\kf_2) 
\,.
\label{splitd2}
\ee
In the following we will often make use of the latter function 
$D_2(\kf_1,\kf_2)$ (without color labels) denoting the 
momentum part of the full $D_2$. 
The amplitude for the production of three gluons in the 
$t$-channel is a superposition of two--gluon amplitudes, 
\be
D_3^{a_1a_2a_3}(\kf_1,\kf_2,\kf_3) 
= \fr{1}{2} g f_{a_1a_2a_3} \,
[ D_2(12,3) - D_2(13,2) + D_2(1,23) ]  
\,. 
\label{d3}
\ee
This is a generalization of the concept of reggeization 
described above for the two--gluon amplitude. 
There we found that the two--gluon amplitude 
in the color octet channel has the analytic properties 
of a one--gluon amplitude, see (\ref{Pole}). In a similar 
way the three gluon amplitude reduces to two--gluon 
amplitudes. In each of the three terms in (\ref{d3}) 
two gluons arrange in such a way that they behave as 
a single gluon. Also here this happens in the color octet 
channel, but we will see later that reggeization can 
also take place in other color channels. 

It can be shown even for arbitrary $n$ that the $n$-gluon 
amplitude $D_n$ contains a part that is a superposition of 
two--gluon amplitudes $D_2$. In the case of the three--gluon 
amplitude this so--called reggeizing part $D_n^R$ exhausts 
the amplitude, whereas for higher $n$ further parts have 
to be added, $D_n=D_n^R+D_n^I$. 
The solution for $D_3$ in (\ref{d3}) is very similar to the expression for 
the quark loop with three gluons attached, cf.\ eq.\ (\ref{d30}). 
It can be obtained from the latter by replacing the lowest 
order amplitude $D_{(2;0)}$ by the full two--gluon 
amplitude $D_2$ while keeping the color structure and 
the momentum arguments. The same procedure allows one to obtain 
the reggeizing part $D_n^R$ from the lowest order coupling 
of $n$ gluons to a quark loop $D_{(n;0)}$ even for arbitrary $n$. 
The quark loop in turn can be computed from the $2^n$ corresponding 
diagrams as described above. For later use we will now give the explicit 
expressions for the reggeizing parts $D_n^R$ of the $n$-gluon amplitudes 
for $n$ up to $5$, 
\bea
\label{d4r}
\lefteqn{D_4^{R\,a_1a_2a_3a_4}(\kf_1,\kf_2,\kf_3,\kf_4)= } \nn \\
  &=&  - g^2 d^{a_1a_2a_3a_4} \, [ D_2(123,4) + D_2(1,234) 
                                  - D_2(14,23) ] 
   \nn \\
 & & - g^2 d^{a_2a_1a_3a_4}  \, [ D_2(134,2) + D_2(124,3)
                                 - D_2(12,34) - D_2(13,24) ] \,,
\eea
\bea
\label{d5r}
\lefteqn{D_5^{R\,a_1a_2a_3a_4a_5}(\kf_1,\kf_2,\kf_3,\kf_4,\kf_5)= } 
%\nn 
\\
  &=&  - g^3 \{ f^{a_1a_2a_3a_4a_5} \, [ 
                  D_2(1234,5) + D_2(1,2345) - D_2(15,234)] 
   \nn \\
  & &  \phantom{ - g^3 } 
              + f^{a_2a_1a_3a_4a_5} \, [ 
                  D_2(1345,2) - D_2(12,345)
                  + D_2(125,34) - D_2(134,25) ]
   \nn \\
  & &  \phantom{ - g^3 } 
              + f^{a_1a_2a_3a_5a_4} \, [ 
                  D_2(1235,4) - D_2(14,235)
                  + D_2(145,23) - D_2(123,45) ]
   \nn \\
  & &  \phantom{ - g^3 } 
              + f^{a_1a_2a_4a_5a_3} \, [ 
                  D_2(1245,3) - D_2(13,245)
                  + D_2(135,24) 
\nn 
%\\
%&& \hspace{3.2cm}
- D_2(124,35) ] \} \,.
\nn
\eea
The explicit expression for $n=6$ can be found in \cite{BE}. 

The remaining parts $D_n^I$ of the $n$-gluon amplitudes 
have been termed `irreducible', because they cannot be reduced 
to two--gluon amplitudes. These parts are known explicitly 
for $n=4$ and $n=5$. In the case of the four--gluon amplitude 
$D_4^I$ has the following structure\footnote{Here $G_4$ and 
$V_{2 \rightarrow 4}$ should be understood as integral operators, 
for the details see again \protect\cite{BE}.} (here we suppress in the 
notation the dependence on the color labels and momenta), 
\be
  D_4^I = G_4 \cdot V_{2 \rightarrow 4} \cdot D_2 = 
\begin{picture}(0,0)%
\includegraphics{solutiond42.pstex}%
\end{picture}%
\setlength{\unitlength}{3947sp}%
\begingroup\makeatletter\ifx\SetFigFont\undefined
% extract first six characters in \fmtname
\def\x#1#2#3#4#5#6#7\relax{\def\x{#1#2#3#4#5#6}}%
\expandafter\x\fmtname xxxxxx\relax \def\y{splain}%
\ifx\x\y   % LaTeX or SliTeX?
\gdef\SetFigFont#1#2#3{%
  \ifnum #1<17\tiny\else \ifnum #1<20\small\else
  \ifnum #1<24\normalsize\else \ifnum #1<29\large\else
  \ifnum #1<34\Large\else \ifnum #1<41\LARGE\else
     \huge\fi\fi\fi\fi\fi\fi
  \csname #3\endcsname}%
\else
\gdef\SetFigFont#1#2#3{\begingroup
  \count@#1\relax \ifnum 25<\count@\count@25\fi
  \def\x{\endgroup\@setsize\SetFigFont{#2pt}}%
  \expandafter\x
    \csname \romannumeral\the\count@ pt\expandafter\endcsname
    \csname @\romannumeral\the\count@ pt\endcsname
  \csname #3\endcsname}%
\fi
\fi\endgroup
\begin{picture}(865,1333)(3691,-2237)
\put(4427,-1861){\makebox(0,0)[lb]{\smash{\SetFigFont{10}{12.0}{rm}$V_{2 \rightarrow 4}$}}}
\end{picture}
 \,.
\label{d4istruct}
\ee
It starts with a two--gluon state coupled to the virtual photons 
via a quark loop (forming together $D_2$). Then there is a 
transition vertex $V_{2 \rightarrow 4}$ coupling the two--gluon 
state to a four--gluon state in the $t$-channel. The vertex is 
known explicitly, but we will not need its explicit form here. 
The four--gluon state $G_4$ is not known analytically so far, but some 
important properties can be derived from its defining integral 
equation. One of these properties is that $D_4^I$ vanishes if one of the 
four gluon momenta vanishes, 
\be
\left. 
  D_4^{I\,a_1a_2a_3a_4}(\kf_1,\kf_2,\kf_3,\kf_4) 
\right|_{\kf_i = 0} = 0  \;\;\; \;\;  (i\in\{1,\dots,4\}) 
\,.
\label{nullstelld4i}
\ee
The `irreducible' part $D_5^I$ of the five--gluon amplitude 
has been found to be a superposition of irreducible four--gluon 
amplitudes $D_4^I$, 
\bea
\label{d5isolution}
%
% PLEASE do not change the layout of this equation.
%
\lefteqn{D_5^{I\,a_1a_2a_3a_4a_5}(\kf_1,\kf_2,\kf_3,\kf_4,\kf_5) =
 \frac{g}{2} \times} \nn \\
&&\times \left\{ f_{a_1a_2c} D_4^{I\,ca_3a_4a_5}(12,3,4,5) 
+ f_{a_1a_3c} D_4^{I\,ca_2a_4a_5}(13,2,4,5) \right. \nn \\
&& \hspace{.5cm} 
+ \,f_{a_1a_4c} D_4^{I\,ca_2a_3a_5}(14,2,3,5) 
+ f_{a_1a_5c} D_4^{I\,ca_2a_3a_4}(15,2,3,4) \nn \\
&& \hspace{.5cm} 
+ \,f_{a_2a_3c} D_4^{I\,a_1ca_4a_5}(1,23,4,5) 
+ f_{a_2a_4c} D_4^{I\,a_1ca_3a_5}(1,24,3,5) \nn \\
&& \hspace{.5cm} 
+\, f_{a_2a_5c} D_4^{I\,a_1ca_3a_4}(1,25,3,4) 
+ f_{a_3a_4c} D_4^{I\,a_1a_2ca_5}(1,2,34,5) \nn \\
&& \hspace{.5cm} \left.
+ \,f_{a_3a_5c} D_4^{I\,a_1a_2ca_4}(1,2,35,4) 
+ f_{a_4a_5c} D_4^{I\,a_1a_2a_3c}(1,2,3,45)
\right\}
\,.
\eea
The fact that $D_5^I$ reduces to a sum of four--gluon 
amplitudes already indicates that the name `irreducible' 
needs more explanation. In fact there is some deeper 
insight hidden here, and we will discuss this issue in detail in 
section \ref{signward}. 
In \cite{BE} an integral equation for the corresponding 
part $D_6^I$ of the six--gluon amplitude has been derived. 
Unfortunately, it has not yet been possible to solve it completely. 
But there are strong indications that the six--gluon amplitude 
consists of a reggeizing part, a part reducing to four--gluon 
amplitudes $D_4^I$ and a third part that contains a full 
six--gluon state, and the corresponding $2$-to-$6$ gluon vertex has 
already been identified. 

\section{Ward type identities}
\label{Ward}

The Ward type identities to be discussed in this section 
relate $n$-gluon amplitudes of different $n$ and allow us to 
gain further insight into the interplay between their color 
and momentum structure. 
These identities of Ward type arise when we 
set one of the $n$ transverse momenta $\kf_i$ to zero. 
Roughly speaking, the amplitude $D_n$ can in this case 
be expressed in terms of the amplitude $D_{n-1}$. 
The reduction is accompanied by an interesting behaviour 
of the corresponding color structure for which we can 
extract a general rule. 
This behavior in color space does not only involve the gluons 
to which the one with vanishing momentum is coupled 
in the original amplitude (for example in the reggeizing 
parts) but involves all gluons of the amplitude. In this sense our 
identities constitute a global property of the amplitudes. 

We will be able to find a formula valid for the 
reggeizing parts $D_n^R$ of the amplitudes and for these 
we can even give a general proof for arbitrary $n$. 
For the remaining parts $D_n^I$ we limit our study to the cases 
$n=4,5$ for which these parts are known explicitly. 
We will then make a conjecture on how the the mechanism 
works for higher $n$ here. If it can be confirmed, 
the identities might provide a valuable tool for the 
further investigation of the field theory structure of 
unitarity corrections. Specifically,  
we will find a characteristic difference between the 
parts of the amplitudes 
that exhibit reggeization and such parts that do not. 
Moreover, the identities lead to strong constraints on the 
amplitudes. 
Both facts might turn out to be very helpful especially for the 
investigation of higher $n$-gluon amplitudes with $n\ge 6$ 
where a more complicated structure is expected to arise. 
We will discuss the potential significance 
of the Ward type identities for the 
field theory structure in more detail in subsection \ref{signward}. 

We start with considering the reggeizing parts $D_n^R$ and 
study for each $n$ how the color tensors rearrange in the case of a 
vanishing momentum $\kf_i$. It seems to us quite instructive to 
see the mechanism at work in concrete examples. 
With these we also hope to convey the impression that the 
Ward type identities impose very strong constraints on the 
color and momentum structure of the amplitudes. 
After that we state the general rule for the amplitudes $D_n^R$ 
in (\ref{ward1}), (\ref{ward2}) and sketch 
the proof for arbitrary $n$. 
Then we turn to the amplitudes $D_4^I$ and $D_5^I$ 
and formulate the conjecture how higher $D_n^I$ have 
to be treated. 

\boldmath
\subsection{The reggeizing parts $D_n^R$}
\unboldmath

After recalling that the BFKL amplitude $D_2$ vanishes if 
one of its momentum arguments vanishes, 
\be
   D_2(\kf_1,\kf_2) |_{\kf_1=0} = D_2 (\kf_1,\kf_2) |_{\kf_2=0} = 0
\,,
\label{ward2repeat}
\label{nullstelld2}
\ee
we consider first the reggeizing parts $D_n^R$ of the amplitudes $D_n$, 
with $n$ ranging from $3$ to $5$. 
This includes also the full amplitude $D_3$ since 
it consists of reggeizing pieces only. 
The simplest relations hold for the case in which we set the first 
momentum $\kf_1 = 0$, namely the vanishing of the 
amplitudes\footnote{Strictly speaking, we here make incorrect use 
of notation since the amplitudes $D_n$ are for different $n$ 
objects in different $\bigotimes_n [su(N_c)]$ tensor spaces.},
\be
  \left. D_3 \right|_{\kf_1=0} 
  = \left. D_4^R \right|_{\kf_1=0} 
  = \left. D_5^R \right|_{\kf_1=0} 
  = 0 \,.
\label{first=0}
\ee
The same is true for setting the $n$th (i.\,e.\ the last) 
momentum to zero in the amplitude $D_n^R$, 
\be
  \left. D_3 \right|_{\kf_3=0} 
  = \left. D_4^R \right|_{\kf_4=0} 
  = \left. D_5^R \right|_{\kf_5=0} 
  = 0 \,.
\label{last=0}
\ee
We will see below that the identities (\ref{first=0}) and 
(\ref{last=0}), although seemingly trivial, fit well into the more 
general rule that determines the color structure of our Ward type 
identities. 
When setting one of the momenta $\kf_2, \dots, \kf_{n-1}$ 
to zero the amplitudes do not vanish. 
For the three--gluon amplitude we find 
\bea
\label{d3k2=0}
  \left. D_3^{a_1a_2a_3} \right|_{\kf_2=0} &=& 
       g f_{a_1a_2a_3} D_2(\kf_1,\kf_3)  \nn \\
%    &=& g f_{a_1a_2c} D_2^{ca_3}(\kf_1,\kf_3)
%       = g f_{ca_2a_3} D_2^{a_1c}(\kf_1,\kf_3) \nn \\
    &=& g \left[ \begin{picture}(0,0)%
\includegraphics{3an2f12.pstex}%
\end{picture}%
\setlength{\unitlength}{3947sp}%
\begingroup\makeatletter\ifx\SetFigFont\undefined
% extract first six characters in \fmtname
\def\x#1#2#3#4#5#6#7\relax{\def\x{#1#2#3#4#5#6}}%
\expandafter\x\fmtname xxxxxx\relax \def\y{splain}%
\ifx\x\y   % LaTeX or SliTeX?
\gdef\SetFigFont#1#2#3{%
  \ifnum #1<17\tiny\else \ifnum #1<20\small\else
  \ifnum #1<24\normalsize\else \ifnum #1<29\large\else
  \ifnum #1<34\Large\else \ifnum #1<41\LARGE\else
     \huge\fi\fi\fi\fi\fi\fi
  \csname #3\endcsname}%
\else
\gdef\SetFigFont#1#2#3{\begingroup
  \count@#1\relax \ifnum 25<\count@\count@25\fi
  \def\x{\endgroup\@setsize\SetFigFont{#2pt}}%
  \expandafter\x
    \csname \romannumeral\the\count@ pt\expandafter\endcsname
    \csname @\romannumeral\the\count@ pt\endcsname
  \csname #3\endcsname}%
\fi
\fi\endgroup
\begin{picture}(249,324)(1714,-973)
\end{picture}
 \right] \star 
              D_2^{b_1b_2}(\kf_1,\kf_3) 
       = g \left[ \,\begin{picture}(0,0)%
\includegraphics{3an2f23.pstex}%
\end{picture}%
\setlength{\unitlength}{3947sp}%
\begingroup\makeatletter\ifx\SetFigFont\undefined
% extract first six characters in \fmtname
\def\x#1#2#3#4#5#6#7\relax{\def\x{#1#2#3#4#5#6}}%
\expandafter\x\fmtname xxxxxx\relax \def\y{splain}%
\ifx\x\y   % LaTeX or SliTeX?
\gdef\SetFigFont#1#2#3{%
  \ifnum #1<17\tiny\else \ifnum #1<20\small\else
  \ifnum #1<24\normalsize\else \ifnum #1<29\large\else
  \ifnum #1<34\Large\else \ifnum #1<41\LARGE\else
     \huge\fi\fi\fi\fi\fi\fi
  \csname #3\endcsname}%
\else
\gdef\SetFigFont#1#2#3{\begingroup
  \count@#1\relax \ifnum 25<\count@\count@25\fi
  \def\x{\endgroup\@setsize\SetFigFont{#2pt}}%
  \expandafter\x
    \csname \romannumeral\the\count@ pt\expandafter\endcsname
    \csname @\romannumeral\the\count@ pt\endcsname
  \csname #3\endcsname}%
\fi
\fi\endgroup
\begin{picture}(249,324)(1639,-973)
\end{picture}
 \right] \star 
              D_2^{b_1b_2}(\kf_1,\kf_3) 
\eea
In the second line we have used the 2--gluon amplitude including 
color labels, that is $D_2^{b_1b_2} = \delta_{b_1b_2} D_2$ 
(see (\ref{splitd2})). 
The way the color structure is written in the second line 
serves to make the general rule for the color structure more 
transparent. Here we have slightly extended the birdtrack 
notation by defining 
\be
  \left[ \begin{picture}(0,0)%
\includegraphics{3an2f12.pstex}%
\end{picture}%
\setlength{\unitlength}{3947sp}%
\begingroup\makeatletter\ifx\SetFigFont\undefined
% extract first six characters in \fmtname
\def\x#1#2#3#4#5#6#7\relax{\def\x{#1#2#3#4#5#6}}%
\expandafter\x\fmtname xxxxxx\relax \def\y{splain}%
\ifx\x\y   % LaTeX or SliTeX?
\gdef\SetFigFont#1#2#3{%
  \ifnum #1<17\tiny\else \ifnum #1<20\small\else
  \ifnum #1<24\normalsize\else \ifnum #1<29\large\else
  \ifnum #1<34\Large\else \ifnum #1<41\LARGE\else
     \huge\fi\fi\fi\fi\fi\fi
  \csname #3\endcsname}%
\else
\gdef\SetFigFont#1#2#3{\begingroup
  \count@#1\relax \ifnum 25<\count@\count@25\fi
  \def\x{\endgroup\@setsize\SetFigFont{#2pt}}%
  \expandafter\x
    \csname \romannumeral\the\count@ pt\expandafter\endcsname
    \csname @\romannumeral\the\count@ pt\endcsname
  \csname #3\endcsname}%
\fi
\fi\endgroup
\begin{picture}(249,324)(1714,-973)
\end{picture}
 \right] \star \Theta^{\{b\}}
   = \begin{picture}(0,0)%
\includegraphics{3an2f12mab.pstex}%
\end{picture}%
\setlength{\unitlength}{3947sp}%
\begingroup\makeatletter\ifx\SetFigFont\undefined
% extract first six characters in \fmtname
\def\x#1#2#3#4#5#6#7\relax{\def\x{#1#2#3#4#5#6}}%
\expandafter\x\fmtname xxxxxx\relax \def\y{splain}%
\ifx\x\y   % LaTeX or SliTeX?
\gdef\SetFigFont#1#2#3{%
  \ifnum #1<17\tiny\else \ifnum #1<20\small\else
  \ifnum #1<24\normalsize\else \ifnum #1<29\large\else
  \ifnum #1<34\Large\else \ifnum #1<41\LARGE\else
     \huge\fi\fi\fi\fi\fi\fi
  \csname #3\endcsname}%
\else
\gdef\SetFigFont#1#2#3{\begingroup
  \count@#1\relax \ifnum 25<\count@\count@25\fi
  \def\x{\endgroup\@setsize\SetFigFont{#2pt}}%
  \expandafter\x
    \csname \romannumeral\the\count@ pt\expandafter\endcsname
    \csname @\romannumeral\the\count@ pt\endcsname
  \csname #3\endcsname}%
\fi
\fi\endgroup
\begin{picture}(249,739)(1714,-1169)
\put(1801,-1111){\makebox(0,0)[lb]{\smash{\SetFigFont{12}{14.4}{rm}{\scriptsize $\{a\}$}}}}
\put(1801,-586){\makebox(0,0)[lb]{\smash{\SetFigFont{12}{14.4}{rm}{\scriptsize $\{b\}$}}}}
\end{picture}
 \star \Theta^{\{b\}}
   = f_{a_1a_2b_1} \delta_{a_3b_2} \Theta^{b_1b_2}
\,,
\ee
where the symbol $\star$ stands for the contraction 
of the set $\{b\}$ of color labels. 
For the reggeizing part $D_4^R$ of the 
four--gluon amplitude we find
\bea
\label{d4rk2=0}
  \left. D_4^{R\,a_1a_2a_3a_4} \right|_{\kf_2=0} &=& 
    g f_{a_1a_2c} D_3^{ca_3a_4}(\kf_1,\kf_3,\kf_4) \\
    &=& g \left[ \begin{picture}(0,0)%
\includegraphics{4an3f12.pstex}%
\end{picture}%
\setlength{\unitlength}{3947sp}%
\begingroup\makeatletter\ifx\SetFigFont\undefined
% extract first six characters in \fmtname
\def\x#1#2#3#4#5#6#7\relax{\def\x{#1#2#3#4#5#6}}%
\expandafter\x\fmtname xxxxxx\relax \def\y{splain}%
\ifx\x\y   % LaTeX or SliTeX?
\gdef\SetFigFont#1#2#3{%
  \ifnum #1<17\tiny\else \ifnum #1<20\small\else
  \ifnum #1<24\normalsize\else \ifnum #1<29\large\else
  \ifnum #1<34\Large\else \ifnum #1<41\LARGE\else
     \huge\fi\fi\fi\fi\fi\fi
  \csname #3\endcsname}%
\else
\gdef\SetFigFont#1#2#3{\begingroup
  \count@#1\relax \ifnum 25<\count@\count@25\fi
  \def\x{\endgroup\@setsize\SetFigFont{#2pt}}%
  \expandafter\x
    \csname \romannumeral\the\count@ pt\expandafter\endcsname
    \csname @\romannumeral\the\count@ pt\endcsname
  \csname #3\endcsname}%
\fi
\fi\endgroup
\begin{picture}(324,324)(1714,-973)
\end{picture}
 \right] \star 
           D_3^{b_1b_2b_3}(\kf_1,\kf_3,\kf_4)  \,.
\eea
This can be seen directly from $D_4^R$. 
For $\kf_2=0$ the two expressions in square brackets in (\ref{d4r}) 
become equal due to (\ref{nullstelld2}) and the color tensor 
in (\ref{d4rk2=0}) is the difference of the two color 
tensors in (\ref{d4r}), 
\be
  d^{abcd} - d^{bacd} = - \fr{1}{2} f_{abk}f_{kcd} \,. 
\label{dabcddbacd}
\ee
The color tensor corresponding to $D_3$, i.\,e.\ 
$f_{b_1b_2b_3}$, is an invariant tensor. According 
to (\ref{invariancecond}) the second line in (\ref{d4rk2=0}) can 
thus also be written as 
\be
  \left. D_4^{R\,a_1a_2a_3a_4} \right|_{\kf_2=0}
  = g \left[\, \begin{picture}(0,0)%
\includegraphics{4an3f23.pstex}%
\end{picture}%
\setlength{\unitlength}{3947sp}%
\begingroup\makeatletter\ifx\SetFigFont\undefined
% extract first six characters in \fmtname
\def\x#1#2#3#4#5#6#7\relax{\def\x{#1#2#3#4#5#6}}%
\expandafter\x\fmtname xxxxxx\relax \def\y{splain}%
\ifx\x\y   % LaTeX or SliTeX?
\gdef\SetFigFont#1#2#3{%
  \ifnum #1<17\tiny\else \ifnum #1<20\small\else
  \ifnum #1<24\normalsize\else \ifnum #1<29\large\else
  \ifnum #1<34\Large\else \ifnum #1<41\LARGE\else
     \huge\fi\fi\fi\fi\fi\fi
  \csname #3\endcsname}%
\else
\gdef\SetFigFont#1#2#3{\begingroup
  \count@#1\relax \ifnum 25<\count@\count@25\fi
  \def\x{\endgroup\@setsize\SetFigFont{#2pt}}%
  \expandafter\x
    \csname \romannumeral\the\count@ pt\expandafter\endcsname
    \csname @\romannumeral\the\count@ pt\endcsname
  \csname #3\endcsname}%
\fi
\fi\endgroup
\begin{picture}(324,324)(1639,-973)
\end{picture}
 
  + \begin{picture}(0,0)%
\includegraphics{4an3f24.pstex}%
\end{picture}%
\setlength{\unitlength}{3947sp}%
\begingroup\makeatletter\ifx\SetFigFont\undefined
% extract first six characters in \fmtname
\def\x#1#2#3#4#5#6#7\relax{\def\x{#1#2#3#4#5#6}}%
\expandafter\x\fmtname xxxxxx\relax \def\y{splain}%
\ifx\x\y   % LaTeX or SliTeX?
\gdef\SetFigFont#1#2#3{%
  \ifnum #1<17\tiny\else \ifnum #1<20\small\else
  \ifnum #1<24\normalsize\else \ifnum #1<29\large\else
  \ifnum #1<34\Large\else \ifnum #1<41\LARGE\else
     \huge\fi\fi\fi\fi\fi\fi
  \csname #3\endcsname}%
\else
\gdef\SetFigFont#1#2#3{\begingroup
  \count@#1\relax \ifnum 25<\count@\count@25\fi
  \def\x{\endgroup\@setsize\SetFigFont{#2pt}}%
  \expandafter\x
    \csname \romannumeral\the\count@ pt\expandafter\endcsname
    \csname @\romannumeral\the\count@ pt\endcsname
  \csname #3\endcsname}%
\fi
\fi\endgroup
\begin{picture}(324,324)(1564,-973)
\end{picture}
\right] \star 
           D_3^{b_1b_2b_3}(\kf_1,\kf_3,\kf_4)  \,.
\ee
For $\kf_3=0$ we find in the same way 
\bea
\label{d4rk3=0}
  \left. D_4^{R\,a_1a_2a_3a_4} \right|_{\kf_3=0} &=&  
    g f_{a_3a_4c} D_3^{a_1a_2c}(\kf_1,\kf_2,\kf_4) \nn \\
   &=& g \left[\, \begin{picture}(0,0)%
\includegraphics{4an3f34.pstex}%
\end{picture}%
\setlength{\unitlength}{3947sp}%
\begingroup\makeatletter\ifx\SetFigFont\undefined
% extract first six characters in \fmtname
\def\x#1#2#3#4#5#6#7\relax{\def\x{#1#2#3#4#5#6}}%
\expandafter\x\fmtname xxxxxx\relax \def\y{splain}%
\ifx\x\y   % LaTeX or SliTeX?
\gdef\SetFigFont#1#2#3{%
  \ifnum #1<17\tiny\else \ifnum #1<20\small\else
  \ifnum #1<24\normalsize\else \ifnum #1<29\large\else
  \ifnum #1<34\Large\else \ifnum #1<41\LARGE\else
     \huge\fi\fi\fi\fi\fi\fi
  \csname #3\endcsname}%
\else
\gdef\SetFigFont#1#2#3{\begingroup
  \count@#1\relax \ifnum 25<\count@\count@25\fi
  \def\x{\endgroup\@setsize\SetFigFont{#2pt}}%
  \expandafter\x
    \csname \romannumeral\the\count@ pt\expandafter\endcsname
    \csname @\romannumeral\the\count@ pt\endcsname
  \csname #3\endcsname}%
\fi
\fi\endgroup
\begin{picture}(324,324)(1564,-973)
\end{picture}
 \right] \star 
           D_3^{b_1b_2b_3}(\kf_1,\kf_2,\kf_4)  \nn \\
   &=& g \left[\, \begin{picture}(0,0)%
\includegraphics{4an3f23.pstex}%
\end{picture}%
\setlength{\unitlength}{3947sp}%
\begingroup\makeatletter\ifx\SetFigFont\undefined
% extract first six characters in \fmtname
\def\x#1#2#3#4#5#6#7\relax{\def\x{#1#2#3#4#5#6}}%
\expandafter\x\fmtname xxxxxx\relax \def\y{splain}%
\ifx\x\y   % LaTeX or SliTeX?
\gdef\SetFigFont#1#2#3{%
  \ifnum #1<17\tiny\else \ifnum #1<20\small\else
  \ifnum #1<24\normalsize\else \ifnum #1<29\large\else
  \ifnum #1<34\Large\else \ifnum #1<41\LARGE\else
     \huge\fi\fi\fi\fi\fi\fi
  \csname #3\endcsname}%
\else
\gdef\SetFigFont#1#2#3{\begingroup
  \count@#1\relax \ifnum 25<\count@\count@25\fi
  \def\x{\endgroup\@setsize\SetFigFont{#2pt}}%
  \expandafter\x
    \csname \romannumeral\the\count@ pt\expandafter\endcsname
    \csname @\romannumeral\the\count@ pt\endcsname
  \csname #3\endcsname}%
\fi
\fi\endgroup
\begin{picture}(324,324)(1639,-973)
\end{picture}
 
          + \begin{picture}(0,0)%
\includegraphics{4an3f13.pstex}%
\end{picture}%
\setlength{\unitlength}{3947sp}%
\begingroup\makeatletter\ifx\SetFigFont\undefined
% extract first six characters in \fmtname
\def\x#1#2#3#4#5#6#7\relax{\def\x{#1#2#3#4#5#6}}%
\expandafter\x\fmtname xxxxxx\relax \def\y{splain}%
\ifx\x\y   % LaTeX or SliTeX?
\gdef\SetFigFont#1#2#3{%
  \ifnum #1<17\tiny\else \ifnum #1<20\small\else
  \ifnum #1<24\normalsize\else \ifnum #1<29\large\else
  \ifnum #1<34\Large\else \ifnum #1<41\LARGE\else
     \huge\fi\fi\fi\fi\fi\fi
  \csname #3\endcsname}%
\else
\gdef\SetFigFont#1#2#3{\begingroup
  \count@#1\relax \ifnum 25<\count@\count@25\fi
  \def\x{\endgroup\@setsize\SetFigFont{#2pt}}%
  \expandafter\x
    \csname \romannumeral\the\count@ pt\expandafter\endcsname
    \csname @\romannumeral\the\count@ pt\endcsname
  \csname #3\endcsname}%
\fi
\fi\endgroup
\begin{picture}(324,324)(1864,-973)
\end{picture}
 \right] \star 
           D_3^{b_1b_2b_3}(\kf_1,\kf_2,\kf_4)  
\,.
\eea
The Ward identities for the reggeizing part $D_5^R$ 
of the five--gluon amplitude arise in a similar way. Setting 
one of the outgoing momenta to zero in (\ref{d5r}), 
we find that the four different momentum structures 
reduce to two. 
When $\kf_2=0$, for instance, the expressions 
in square brackets in line 1 and 2 in (\ref{d5r}) 
become equal up to a sign, as do the expressions in square brackets 
in line 3 and 4. The corresponding pairs of color tensors 
can be added using (\ref{algebra}), 
\bea
 f^{a_1a_2a_3a_4a_5} - f^{a_2a_1a_3a_4a_5} &=& 
  f_{a_1a_2c}\,d^{ca_3a_4a_5}
\label{fminusf1}
\\
f^{a_1a_2a_3a_5a_4} -  f^{a_1a_2a_4a_5a_3} &=& 
  f_{a_1a_2c}\,d^{a_3ca_4a_5}
\label{fminusf2}
\,.
\eea
Comparing with (\ref{d4r}) we can thus write 
\bea
\label{d5rk2=0} 
\left. D_5^{R\,a_1a_2a_3a_4a_5} \right|_{\kf_2=0} 
\hspace{-.5cm} &&=
  g f_{a_1a_2c} D_4^{R\,ca_3a_4a_5}(\kf_1,\kf_3,\kf_4,\kf_5) \nn \\
  &&= g \left[ \begin{picture}(0,0)%
\includegraphics{5an4f12.pstex}%
\end{picture}%
\setlength{\unitlength}{3947sp}%
\begingroup\makeatletter\ifx\SetFigFont\undefined
% extract first six characters in \fmtname
\def\x#1#2#3#4#5#6#7\relax{\def\x{#1#2#3#4#5#6}}%
\expandafter\x\fmtname xxxxxx\relax \def\y{splain}%
\ifx\x\y   % LaTeX or SliTeX?
\gdef\SetFigFont#1#2#3{%
  \ifnum #1<17\tiny\else \ifnum #1<20\small\else
  \ifnum #1<24\normalsize\else \ifnum #1<29\large\else
  \ifnum #1<34\Large\else \ifnum #1<41\LARGE\else
     \huge\fi\fi\fi\fi\fi\fi
  \csname #3\endcsname}%
\else
\gdef\SetFigFont#1#2#3{\begingroup
  \count@#1\relax \ifnum 25<\count@\count@25\fi
  \def\x{\endgroup\@setsize\SetFigFont{#2pt}}%
  \expandafter\x
    \csname \romannumeral\the\count@ pt\expandafter\endcsname
    \csname @\romannumeral\the\count@ pt\endcsname
  \csname #3\endcsname}%
\fi
\fi\endgroup
\begin{picture}(399,324)(1714,-973)
\end{picture}
 \right] \star 
     D_4^{R\,b_1b_2b_3b_4}(\kf_1,\kf_3,\kf_4,\kf_5)  \nn \\
&&= g \left[\, \begin{picture}(0,0)%
\includegraphics{5an4f23.pstex}%
\end{picture}%
\setlength{\unitlength}{3947sp}%
\begingroup\makeatletter\ifx\SetFigFont\undefined
% extract first six characters in \fmtname
\def\x#1#2#3#4#5#6#7\relax{\def\x{#1#2#3#4#5#6}}%
\expandafter\x\fmtname xxxxxx\relax \def\y{splain}%
\ifx\x\y   % LaTeX or SliTeX?
\gdef\SetFigFont#1#2#3{%
  \ifnum #1<17\tiny\else \ifnum #1<20\small\else
  \ifnum #1<24\normalsize\else \ifnum #1<29\large\else
  \ifnum #1<34\Large\else \ifnum #1<41\LARGE\else
     \huge\fi\fi\fi\fi\fi\fi
  \csname #3\endcsname}%
\else
\gdef\SetFigFont#1#2#3{\begingroup
  \count@#1\relax \ifnum 25<\count@\count@25\fi
  \def\x{\endgroup\@setsize\SetFigFont{#2pt}}%
  \expandafter\x
    \csname \romannumeral\the\count@ pt\expandafter\endcsname
    \csname @\romannumeral\the\count@ pt\endcsname
  \csname #3\endcsname}%
\fi
\fi\endgroup
\begin{picture}(399,324)(1639,-973)
\end{picture}
 + \begin{picture}(0,0)%
\includegraphics{5an4f24.pstex}%
\end{picture}%
\setlength{\unitlength}{3947sp}%
\begingroup\makeatletter\ifx\SetFigFont\undefined
% extract first six characters in \fmtname
\def\x#1#2#3#4#5#6#7\relax{\def\x{#1#2#3#4#5#6}}%
\expandafter\x\fmtname xxxxxx\relax \def\y{splain}%
\ifx\x\y   % LaTeX or SliTeX?
\gdef\SetFigFont#1#2#3{%
  \ifnum #1<17\tiny\else \ifnum #1<20\small\else
  \ifnum #1<24\normalsize\else \ifnum #1<29\large\else
  \ifnum #1<34\Large\else \ifnum #1<41\LARGE\else
     \huge\fi\fi\fi\fi\fi\fi
  \csname #3\endcsname}%
\else
\gdef\SetFigFont#1#2#3{\begingroup
  \count@#1\relax \ifnum 25<\count@\count@25\fi
  \def\x{\endgroup\@setsize\SetFigFont{#2pt}}%
  \expandafter\x
    \csname \romannumeral\the\count@ pt\expandafter\endcsname
    \csname @\romannumeral\the\count@ pt\endcsname
  \csname #3\endcsname}%
\fi
\fi\endgroup
\begin{picture}(399,324)(1564,-973)
\end{picture}
 
  +  \begin{picture}(0,0)%
\includegraphics{5an4f25r.pstex}%
\end{picture}%
\setlength{\unitlength}{3947sp}%
\begingroup\makeatletter\ifx\SetFigFont\undefined
% extract first six characters in \fmtname
\def\x#1#2#3#4#5#6#7\relax{\def\x{#1#2#3#4#5#6}}%
\expandafter\x\fmtname xxxxxx\relax \def\y{splain}%
\ifx\x\y   % LaTeX or SliTeX?
\gdef\SetFigFont#1#2#3{%
  \ifnum #1<17\tiny\else \ifnum #1<20\small\else
  \ifnum #1<24\normalsize\else \ifnum #1<29\large\else
  \ifnum #1<34\Large\else \ifnum #1<41\LARGE\else
     \huge\fi\fi\fi\fi\fi\fi
  \csname #3\endcsname}%
\else
\gdef\SetFigFont#1#2#3{\begingroup
  \count@#1\relax \ifnum 25<\count@\count@25\fi
  \def\x{\endgroup\@setsize\SetFigFont{#2pt}}%
  \expandafter\x
    \csname \romannumeral\the\count@ pt\expandafter\endcsname
    \csname @\romannumeral\the\count@ pt\endcsname
  \csname #3\endcsname}%
\fi
\fi\endgroup
\begin{picture}(399,324)(1639,-973)
\end{picture}
 \right] \star 
     D_4^{R\,b_1b_2b_3b_4}(\kf_1,\kf_3,\kf_4,\kf_5) 
\,.
\eea
For $\kf_4=0$ similarly 
\bea
\label{d5rk4=0}
  \left. D_5^{R\,a_1a_2a_3a_4a_5} \right|_{\kf_4=0} 
\hspace{-.5cm} &&=
  g f_{a_4a_5c} D_4^{R\,a_1a_2a_3c}(\kf_1,\kf_2,\kf_3,\kf_5) \nn \\
  &&= g \left[\, \begin{picture}(0,0)%
\includegraphics{5an4f45.pstex}%
\end{picture}%
\setlength{\unitlength}{3947sp}%
\begingroup\makeatletter\ifx\SetFigFont\undefined
% extract first six characters in \fmtname
\def\x#1#2#3#4#5#6#7\relax{\def\x{#1#2#3#4#5#6}}%
\expandafter\x\fmtname xxxxxx\relax \def\y{splain}%
\ifx\x\y   % LaTeX or SliTeX?
\gdef\SetFigFont#1#2#3{%
  \ifnum #1<17\tiny\else \ifnum #1<20\small\else
  \ifnum #1<24\normalsize\else \ifnum #1<29\large\else
  \ifnum #1<34\Large\else \ifnum #1<41\LARGE\else
     \huge\fi\fi\fi\fi\fi\fi
  \csname #3\endcsname}%
\else
\gdef\SetFigFont#1#2#3{\begingroup
  \count@#1\relax \ifnum 25<\count@\count@25\fi
  \def\x{\endgroup\@setsize\SetFigFont{#2pt}}%
  \expandafter\x
    \csname \romannumeral\the\count@ pt\expandafter\endcsname
    \csname @\romannumeral\the\count@ pt\endcsname
  \csname #3\endcsname}%
\fi
\fi\endgroup
\begin{picture}(399,324)(1489,-973)
\end{picture}
 \right] \star 
  D_4^{R\,b_1b_2b_3b_4}(\kf_1,\kf_2,\kf_3,\kf_5) \nn \\
  &&= g \left[\, \begin{picture}(0,0)%
\includegraphics{5an4f34.pstex}%
\end{picture}%
\setlength{\unitlength}{3947sp}%
\begingroup\makeatletter\ifx\SetFigFont\undefined
% extract first six characters in \fmtname
\def\x#1#2#3#4#5#6#7\relax{\def\x{#1#2#3#4#5#6}}%
\expandafter\x\fmtname xxxxxx\relax \def\y{splain}%
\ifx\x\y   % LaTeX or SliTeX?
\gdef\SetFigFont#1#2#3{%
  \ifnum #1<17\tiny\else \ifnum #1<20\small\else
  \ifnum #1<24\normalsize\else \ifnum #1<29\large\else
  \ifnum #1<34\Large\else \ifnum #1<41\LARGE\else
     \huge\fi\fi\fi\fi\fi\fi
  \csname #3\endcsname}%
\else
\gdef\SetFigFont#1#2#3{\begingroup
  \count@#1\relax \ifnum 25<\count@\count@25\fi
  \def\x{\endgroup\@setsize\SetFigFont{#2pt}}%
  \expandafter\x
    \csname \romannumeral\the\count@ pt\expandafter\endcsname
    \csname @\romannumeral\the\count@ pt\endcsname
  \csname #3\endcsname}%
\fi
\fi\endgroup
\begin{picture}(399,324)(1564,-973)
\end{picture}
 +\begin{picture}(0,0)%
\includegraphics{5an4f24l.pstex}%
\end{picture}%
\setlength{\unitlength}{3947sp}%
\begingroup\makeatletter\ifx\SetFigFont\undefined
% extract first six characters in \fmtname
\def\x#1#2#3#4#5#6#7\relax{\def\x{#1#2#3#4#5#6}}%
\expandafter\x\fmtname xxxxxx\relax \def\y{splain}%
\ifx\x\y   % LaTeX or SliTeX?
\gdef\SetFigFont#1#2#3{%
  \ifnum #1<17\tiny\else \ifnum #1<20\small\else
  \ifnum #1<24\normalsize\else \ifnum #1<29\large\else
  \ifnum #1<34\Large\else \ifnum #1<41\LARGE\else
     \huge\fi\fi\fi\fi\fi\fi
  \csname #3\endcsname}%
\else
\gdef\SetFigFont#1#2#3{\begingroup
  \count@#1\relax \ifnum 25<\count@\count@25\fi
  \def\x{\endgroup\@setsize\SetFigFont{#2pt}}%
  \expandafter\x
    \csname \romannumeral\the\count@ pt\expandafter\endcsname
    \csname @\romannumeral\the\count@ pt\endcsname
  \csname #3\endcsname}%
\fi
\fi\endgroup
\begin{picture}(399,324)(1189,-973)
\end{picture}
  
  + \begin{picture}(0,0)%
\includegraphics{5an4f14l.pstex}%
\end{picture}%
\setlength{\unitlength}{3947sp}%
\begingroup\makeatletter\ifx\SetFigFont\undefined
% extract first six characters in \fmtname
\def\x#1#2#3#4#5#6#7\relax{\def\x{#1#2#3#4#5#6}}%
\expandafter\x\fmtname xxxxxx\relax \def\y{splain}%
\ifx\x\y   % LaTeX or SliTeX?
\gdef\SetFigFont#1#2#3{%
  \ifnum #1<17\tiny\else \ifnum #1<20\small\else
  \ifnum #1<24\normalsize\else \ifnum #1<29\large\else
  \ifnum #1<34\Large\else \ifnum #1<41\LARGE\else
     \huge\fi\fi\fi\fi\fi\fi
  \csname #3\endcsname}%
\else
\gdef\SetFigFont#1#2#3{\begingroup
  \count@#1\relax \ifnum 25<\count@\count@25\fi
  \def\x{\endgroup\@setsize\SetFigFont{#2pt}}%
  \expandafter\x
    \csname \romannumeral\the\count@ pt\expandafter\endcsname
    \csname @\romannumeral\the\count@ pt\endcsname
  \csname #3\endcsname}%
\fi
\fi\endgroup
\begin{picture}(399,324)(1264,-973)
\end{picture}
 \right] \star 
  D_4^{R\,b_1b_2b_3b_4}(\kf_1,\kf_2,\kf_3,\kf_5)
\,.
\eea
The last lines in (\ref{d5rk2=0}) and (\ref{d5rk4=0}) are again 
implied by the fact that the color tensors in $D_4^{R\,b_1b_2b_3b_4}$ 
are invariant tensors. 
For $\kf_3=0$ we need the analogue of (\ref{fminusf1}), 
(\ref{fminusf2}) which is slightly more complicated. 
Applying (\ref{algebra}) two times we get the two identities 
\bea
  f^{a_1a_2a_3a_4a_5} - f^{a_1a_2a_4a_5a_3} &=& 
   f_{a_1a_3c}\,d^{ca_2a_4a_5} + f_{a_2a_3c}\,d^{a_1ca_4a_5}
\\
  f^{a_2a_1a_3a_4a_5} + f^{a_1a_2a_3a_5a_4} &=&
   f_{a_1a_3c}\,d^{a_2ca_4a_5} + f_{a_2a_3c}\,d^{ca_1a_4a_5} 
\,.
\eea
Using this we get from the formula (\ref{d5r}) for the 
reggeizing part 
\bea
\label{d5rk3=0}
  \left. D_5^{R\,a_1a_2a_3a_4a_5} \right|_{\kf_3=0} &=&
  g f_{a_1a_3c} D_4^{R\,ca_2a_4a_5}(\kf_1,\kf_2,\kf_4,\kf_5) 
  + g f_{a_2a_3c} D_4^{R\,a_1ca_4a_5}(\kf_1,\kf_2,\kf_4,\kf_5) \nn \\
  &=& g \left[\, \begin{picture}(0,0)%
\includegraphics{5an4f13.pstex}%
\end{picture}%
\setlength{\unitlength}{3947sp}%
\begingroup\makeatletter\ifx\SetFigFont\undefined
% extract first six characters in \fmtname
\def\x#1#2#3#4#5#6#7\relax{\def\x{#1#2#3#4#5#6}}%
\expandafter\x\fmtname xxxxxx\relax \def\y{splain}%
\ifx\x\y   % LaTeX or SliTeX?
\gdef\SetFigFont#1#2#3{%
  \ifnum #1<17\tiny\else \ifnum #1<20\small\else
  \ifnum #1<24\normalsize\else \ifnum #1<29\large\else
  \ifnum #1<34\Large\else \ifnum #1<41\LARGE\else
     \huge\fi\fi\fi\fi\fi\fi
  \csname #3\endcsname}%
\else
\gdef\SetFigFont#1#2#3{\begingroup
  \count@#1\relax \ifnum 25<\count@\count@25\fi
  \def\x{\endgroup\@setsize\SetFigFont{#2pt}}%
  \expandafter\x
    \csname \romannumeral\the\count@ pt\expandafter\endcsname
    \csname @\romannumeral\the\count@ pt\endcsname
  \csname #3\endcsname}%
\fi
\fi\endgroup
\begin{picture}(399,324)(1864,-973)
\end{picture}
 
           + \begin{picture}(0,0)%
\includegraphics{5an4f23.pstex}%
\end{picture}%
\setlength{\unitlength}{3947sp}%
\begingroup\makeatletter\ifx\SetFigFont\undefined
% extract first six characters in \fmtname
\def\x#1#2#3#4#5#6#7\relax{\def\x{#1#2#3#4#5#6}}%
\expandafter\x\fmtname xxxxxx\relax \def\y{splain}%
\ifx\x\y   % LaTeX or SliTeX?
\gdef\SetFigFont#1#2#3{%
  \ifnum #1<17\tiny\else \ifnum #1<20\small\else
  \ifnum #1<24\normalsize\else \ifnum #1<29\large\else
  \ifnum #1<34\Large\else \ifnum #1<41\LARGE\else
     \huge\fi\fi\fi\fi\fi\fi
  \csname #3\endcsname}%
\else
\gdef\SetFigFont#1#2#3{\begingroup
  \count@#1\relax \ifnum 25<\count@\count@25\fi
  \def\x{\endgroup\@setsize\SetFigFont{#2pt}}%
  \expandafter\x
    \csname \romannumeral\the\count@ pt\expandafter\endcsname
    \csname @\romannumeral\the\count@ pt\endcsname
  \csname #3\endcsname}%
\fi
\fi\endgroup
\begin{picture}(399,324)(1639,-973)
\end{picture}
\right] \star 
  D_4^{R\,b_1b_2b_3b_4}(\kf_1,\kf_2,\kf_4,\kf_5) \nn \\
  &=& g \left[\, \begin{picture}(0,0)%
\includegraphics{5an4f34.pstex}%
\end{picture}%
\setlength{\unitlength}{3947sp}%
\begingroup\makeatletter\ifx\SetFigFont\undefined
% extract first six characters in \fmtname
\def\x#1#2#3#4#5#6#7\relax{\def\x{#1#2#3#4#5#6}}%
\expandafter\x\fmtname xxxxxx\relax \def\y{splain}%
\ifx\x\y   % LaTeX or SliTeX?
\gdef\SetFigFont#1#2#3{%
  \ifnum #1<17\tiny\else \ifnum #1<20\small\else
  \ifnum #1<24\normalsize\else \ifnum #1<29\large\else
  \ifnum #1<34\Large\else \ifnum #1<41\LARGE\else
     \huge\fi\fi\fi\fi\fi\fi
  \csname #3\endcsname}%
\else
\gdef\SetFigFont#1#2#3{\begingroup
  \count@#1\relax \ifnum 25<\count@\count@25\fi
  \def\x{\endgroup\@setsize\SetFigFont{#2pt}}%
  \expandafter\x
    \csname \romannumeral\the\count@ pt\expandafter\endcsname
    \csname @\romannumeral\the\count@ pt\endcsname
  \csname #3\endcsname}%
\fi
\fi\endgroup
\begin{picture}(399,324)(1564,-973)
\end{picture}
 
           + \begin{picture}(0,0)%
\includegraphics{5an4f35.pstex}%
\end{picture}%
\setlength{\unitlength}{3947sp}%
\begingroup\makeatletter\ifx\SetFigFont\undefined
% extract first six characters in \fmtname
\def\x#1#2#3#4#5#6#7\relax{\def\x{#1#2#3#4#5#6}}%
\expandafter\x\fmtname xxxxxx\relax \def\y{splain}%
\ifx\x\y   % LaTeX or SliTeX?
\gdef\SetFigFont#1#2#3{%
  \ifnum #1<17\tiny\else \ifnum #1<20\small\else
  \ifnum #1<24\normalsize\else \ifnum #1<29\large\else
  \ifnum #1<34\Large\else \ifnum #1<41\LARGE\else
     \huge\fi\fi\fi\fi\fi\fi
  \csname #3\endcsname}%
\else
\gdef\SetFigFont#1#2#3{\begingroup
  \count@#1\relax \ifnum 25<\count@\count@25\fi
  \def\x{\endgroup\@setsize\SetFigFont{#2pt}}%
  \expandafter\x
    \csname \romannumeral\the\count@ pt\expandafter\endcsname
    \csname @\romannumeral\the\count@ pt\endcsname
  \csname #3\endcsname}%
\fi
\fi\endgroup
\begin{picture}(399,324)(1489,-973)
\end{picture}
\right] \star 
  D_4^{R\,b_1b_2b_3b_4}(\kf_1,\kf_2,\kf_4,\kf_5) 
\,.
\eea

The Ward identities collected here for the reggeizing parts $D_n^R$ 
of the $n$-gluon amplitudes ($n\ge 3$) can be summarized as 
follows. For vanishing momentum $\kf_i$ the momentum 
part of the amplitude $D_n^R$ reduces to $D_{n-1}^R$, 
the momentum arguments being the $(n-1)$ remaining 
transverse momenta. 
(Here we again identify $D_3^R=D_3$ since $D_3$ reggeizes 
completely, and $D_{3-1}^R$ should be understood as $D_2$.) 
In color space the label $a_i$ of the zero--momentum gluon 
has to be contracted via a $f_{a_ja_ic}$ tensor with the color labels 
of all gluons $j$ to the left, that is with $j<i$, and these contractions 
have to be added. The label $c$ has to be taken at the $j$th position 
in the amplitude $D_{n-1}^R$. 
Since the amplitudes $D_{n-1}^R$ consist of 
invariant tensors in color space, we can alternatively contract the 
label $a_i$ with all color labels $a_j$ to the right ($j>i$) with 
a $f_{a_ia_jc}$ tensor. In the latter case the label $c$ is at 
the $(j-1)$th position in the amplitude $D_{n-1}^R$. 
Casting this into formulae we have 
\bea
\lefteqn{
\left. D_n^{R\,a_1\dots a_n} (\kf_1, \dots, \kf_n) \right|_{\kf_i=0} = 
} \nn \\
&&= g \sum_{j=1}^{i-1} f_{a_ja_ic} 
       D_{n-1\hspace{.57cm}\uparrow{\mbox{\tiny $j$}}}^{R\,a_1 
      \dots \hat{a}_jc \dots \hat{a}_i \dots a_n} 
       (\kf_1, \dots, \widehat{\kf}_i, \dots, \kf_n) 
 \label{ward1}\\ 
 &&= g \sum_{j=i+1}^n f_{a_ia_jc} 
       D_{n-1\hspace{.8cm}\uparrow{\mbox{\tiny $j\!-\!1$}}}^{R\,a_1 
     \dots \hat{a}_i \dots c \hat{a}_j  \dots a_n}
       (\kf_1, \dots, \widehat{\kf}_i, \dots, \kf_n) 
\,,
\label{ward2}
\eea
where the hat indicates that the corresponding quantity has to be left out. 
The formulae include the special cases 
$\kf_1=0$ and $\kf_n=0$ as well: the respective sum 
in (\ref{ward1}) or (\ref{ward2}) is empty or it contains $(n-1)$ 
terms and vanishes due to 
the condition for invariant tensors (\ref{invariancecond}). 

In \cite{BE} also the reggeizing part $D_6^R$ of the six--gluon 
has been calculated explicitly. 
It can also be shown to fulfill the Ward type identities (\ref{ward1}), 
(\ref{ward2}) following the same lines as for up to five gluons. 

These Ward type identities for the reggeizing part of the $n$-gluon 
amplitude can even be shown to hold for arbitrary $n$. 
The reason for this is that they are 
obtained from the corresponding quark loop $D_{(n;0)}$ by 
replacing $D_{(2;0)} \to D_2$ as discussed in section 
\ref{reviewmultigluon}. 
Due to this construction 
it is sufficient to prove the identities for $D_{(n;0)}$. 
The quark loop is the sum of $2^n$ diagrams
%\footnote{As 
%mentioned already in section \protect\ref{reviewmultigluon} 
%the two diagrams with all gluons coupled to the same line 
%(quark or antiquark) do not give rise to a separate $D_2$ term 
%in the reggeizing parts since in the quark loop they act as 
%regularization terms only. We can therefore disregard them 
%for the present discussion.}. 
Consider now two of these diagrams that 
differ in the coupling of the $i$th gluon. It is coupled to the 
quark line in one and to the antiquark line in the other diagram. 
Due to this the two diagrams have opposite sign. But otherwise 
the momentum structure is the same when setting $\kf_i=0$. 
The color structure differs by the position of the generator 
$t^{a_i}$ within the trace of generators around the loop. 
Starting from one of the two diagrams 
we can shift $t^{a_i}$ around the loop to the left (or to the right) 
by iterated use of (\ref{algebra}). Doing so we come across 
all gluons $j$ with $j<i$ since the cuts in the amplitude 
(see Fig.\ \ref{fig:nocrossing}) forbid crossing of $t$-channel 
gluons. Therefore this procedure generates 
exactly the terms needed for (\ref{ward1}) containing a trace 
over $(n-1)$ generators contracted with a $f_{a_ja_ic}$. 
(Although it is a bit tedious, the correct signs can be checked 
without difficulty.) The two remaining 
terms with a trace over $n$ generators cancel due to their 
different sign mentioned above. The second form of the Ward type identity 
(\ref{ward2}) is obtained by shifting the generator $t^{a_i}$ around 
the loop to the right instead of to the left. 

\boldmath
\subsection{The amplitudes $D_4^I$ and $D_5^I$}
\unboldmath

We now come to examining the Ward type identities for that 
part of the $n$-gluon amplitude which is not the superposition 
of two--gluon amplitudes. This requires $n\ge 4$ since only in this 
case we have a non-vanishing part $D_n^I$. 
On the other hand our knowledge of this part is rather 
limited: it is only up to $n=5$ that we know its structure, 
and even there we do not have an analytic formula for 
the four--gluon state. 
Therefore we have to restrict ourselves 
to $D_4^I$ and $D_5^I$ here. 

For $D_4^I$ we already know 
(see (\ref{nullstelld4i})) that it vanishes whenever 
one of the momentum arguments vanishes. 
The study of the amplitude $D_5^I$ is surprisingly simple. 
It fulfills Ward type identities very similar to those valid for 
the reggeizing amplitudes $D_n^R$, i.\,e.\ (\ref{ward1}) 
and (\ref{ward2}). The only difference is well in agreement with 
what one would naturally expect. Whereas the amplitude 
$D_n^R$ was reduced to a superposition of $D_{n-1}^R$'s 
when one momentum was set to zero, 
$D_5^I$ now reduces to a sum of irreducible $D_4^I$ amplitudes. 
In detail, we find 
\bea
\lefteqn{
\left. D_5^{I\,a_1\dots a_5} (\kf_1, \dots, \kf_5) \right|_{\kf_i=0} =
}\nn \\
 &&= g \sum_{j=1}^{i-1} f_{a_ja_ic} 
       D_4^{I\,a_1 \dots \hat{a}_jc \dots \hat{a}_i \dots a_5} 
       (\kf_1, \dots, \widehat{\kf}_i, \dots, \kf_5) 
 \label{d5iward1}\\ 
 &&= g \sum_{j=i+1}^5 f_{a_ia_jc} 
       D_4^{I\,a_1 \dots \hat{a}_i \dots c \hat{a}_j  \dots a_n}
       (\kf_1, \dots, \widehat{\kf}_i, \dots, \kf_5) 
\label{d5iward2}
\eea
for any $i \in \{ 1,\dots,5 \} $. 
Once we have found the amplitude $D_5^I$ 
in the form (\ref{d5isolution}), only two more ingredients 
are required for the proof. One is the vanishing of $D_4^I$ for 
vanishing argument (\ref{nullstelld4i}), the other is the 
defining property (\ref{invariancecond}) of 
invariant $\mbox{su}(N_c)$ tensors. 
The latter applies here since the four gluons in the irreducible 
amplitude $D_4^I$ are in a color singlet state. With these two 
pieces of information at hand the calculation leading to 
(\ref{d5iward1}) and (\ref{d5iward2}) is almost trivial. 

\subsection{Significance of the Ward type identities}
\label{signward}

The Ward type identities discussed so far suggest an underlying pattern 
also valid for higher $n$-gluon amplitudes. 
In this section we try to make several conjectures concerning 
this pattern for higher $n$. 
Each of the conjectures applies to a certain part of the 
amplitudes, for instance to the part of a $n$-gluon amplitude 
that reggeizes into two--gluon amplitudes. 
Since the original integral equations only constrain the full amplitudes 
$D_n$ we will not be able to prove the conjectures for the 
different parts separately. The main conjecture we make here 
is that it is in fact possible to split the amplitudes into 
different parts in such a way that those parts 
fulfill the conjectures separately. 
This main conjecture can of course be tested 
by investigating amplitudes with more $t$-channel gluons 
or by deriving the field theory structure from a different starting 
point. 

Our first conjecture 
is that the number--changing vertices of the effective field theory 
should vanish when one of the outgoing gluons has vanishing 
transverse momentum. This is true \cite{BW} 
for the two--to--four transition vertex 
$V_{2 \rarr 4}$ as well as for the two--to--six gluon transition 
vertex derived in \cite{BE}. 

Connected with the preceding one is the conjecture 
that the non--reggeizing parts of the amplitudes, i.\,e.\ 
the parts that cannot be written as superpositions of 
lower amplitudes,  
vanish if one of the outgoing gluons has zero momentum. 
Examples found so far are the two--gluon (BFKL) amplitude $D_2$ 
and the irreducible part $D_4^I$ of the four--gluon amplitude. 

The final and probably the most important 
conjecture concerns the reggeizing parts of the amplitudes. 
If a part of a $n$-gluon amplitude can be written as a superposition 
of non--reggeizing parts of lower amplitudes 
it should satisfy the Ward type identities found in the previous sections. 

In order to understand the significance of these conjectures 
it will be useful to discuss the issue of decomposing the 
amplitudes into the two parts $D_n^R$ and $D_n^I$. 
Initially, the study of the $n$-gluon amplitudes starts with a 
set of integral equations for the full amplitudes. 
The first step is then to {\sl choose} a reggeizing part and 
to derive a new integral equation for the remaining part. 
It has turned out that the choice described in section 
\ref{reviewmultigluon}, namely a superposition of two--gluon 
amplitudes constructed from the quark loop with $n$ gluons, 
is singled out because it leads to a particularly simple 
equation for the remaining part. Only due to this it was 
possible to discover the field theory structure in the amplitudes. 
Other choices for the reggeizing part would lead 
to very complicated equations for the remaining part, 
and it will in general be difficult to learn something from them. 
From this point of view our 
conjectures about the Ward type identities can be understood 
as conditions for the sensible decomposition of the amplitudes 
into its different parts which then leads to the identification the 
different elements of the effective field theory. 

We expect such conditions to become especially helpful already in the 
the course of investigating the six--gluon amplitude. 
To illustrate the potential significance of the Ward type identities 
let us briefly discuss the six--gluon amplitude $D_6$. 
In the first step, the quark loop offers sufficient inspiration for 
the sensible choice of a reggeizing part. However, 
in the six--gluon amplitude a new problem arises. 
After decomposing the amplitude in the canonical way into 
reggeizing part and a remaining part, $D_6 = D_6^R + D_6^I$, 
a new integral equation for $D_6^I$ was found \cite{BE}. 
From its structure it is obvious that 
a further decomposition is required to fully understand 
its structure. Namely, the part 
$D_6^I$ will must contain a part that is the superposition 
of irreducible four--gluon amplitudes $D_4^I$, symbolically 
\be
  D_6^I = D_6^{I,\,R} + D_6^{I,\,I} = \sum D_4^I + D_6^{I,\,I} \,. 
\ee
This means that the part $D_6^{I,\,R}$ is irreducible with 
respect to the two--gluon amplitude, but it is reducible with 
respect to the four--gluon amplitude. The part $D_6^{I,\,I}$ 
is irreducible with respect to both the two--gluon and the 
four--gluon amplitude. 
But in this case we do not have a quark loop suggesting a 
good choice for $D_6^{I,\,R}$. Exactly at this point  
the Ward type identities will be very useful for 
identifying a correct choice for the reggeizing part in this 
decomposition. 
To summarize, we expect roughly the following structure 
to arise in higher $n$-gluon amplitudes. 
There will be irreducible 
$m$-gluon compound states for all even $m$. 
Based on each of them there will be a hierarchy of 
reggeizing parts of amplitudes, 
all of them reggeizing with respect to 
the same $m$-gluon compound state. The amplitudes 
in each of these hierarchies should then 
obey Ward type identities of the kind discussed above. 

\section{Reggeization tensors}
\label{tensors}

The preceding section was devoted to the study of the 
more global interplay of color and momentum structure in 
the $n$-gluon amplitudes. 
Now we turn to more local properties of the color structure. 
Namely, we will be able to assign to a reggeized gluon a kind 
of 'quantum number' that specifies its behavior in the process 
of reggeization. 
The cleanest environment for studying the 
mechanism of reggeization is clearly provided by the two--reggeon 
compound state or BFKL amplitude. 
To see how higher and higher 'Fock states' of the reggeized gluon 
occur we will therefore investigate the reggeizing parts $D_n^R$ of 
the $n$-gluon amplitudes that consist of superpositions of two--gluon 
amplitudes. 
Here we will focus on the color structure of single terms. 
We will mainly deal with color algebra, and 
one should be careful in drawing conclusions which go 
substantially beyond the subject of color structure in this context. 
At the end we will of course try to interpret the 
results in a larger context. 

The process of reggeization can be viewed 
in two different ways. To illustrate this let us have a look at the 
reggeizing part $D_4^R$ of the four--gluon amplitude (\ref{d4r}). 
It can be represented diagrammatically as
\be
  D_4^R(\kf_1,\kf_2,\kf_3,\kf_4) 
 = \sum \begin{picture}(0,0)%
\includegraphics{solutiond41.pstex}%
\end{picture}%
\setlength{\unitlength}{3947sp}%
\begingroup\makeatletter\ifx\SetFigFont\undefined
% extract first six characters in \fmtname
\def\x#1#2#3#4#5#6#7\relax{\def\x{#1#2#3#4#5#6}}%
\expandafter\x\fmtname xxxxxx\relax \def\y{splain}%
\ifx\x\y   % LaTeX or SliTeX?
\gdef\SetFigFont#1#2#3{%
  \ifnum #1<17\tiny\else \ifnum #1<20\small\else
  \ifnum #1<24\normalsize\else \ifnum #1<29\large\else
  \ifnum #1<34\Large\else \ifnum #1<41\LARGE\else
     \huge\fi\fi\fi\fi\fi\fi
  \csname #3\endcsname}%
\else
\gdef\SetFigFont#1#2#3{\begingroup
  \count@#1\relax \ifnum 25<\count@\count@25\fi
  \def\x{\endgroup\@setsize\SetFigFont{#2pt}}%
  \expandafter\x
    \csname \romannumeral\the\count@ pt\expandafter\endcsname
    \csname @\romannumeral\the\count@ pt\endcsname
  \csname #3\endcsname}%
\fi
\fi\endgroup
\begin{picture}(865,966)(2373,-2016)
\end{picture}
 
\,.
\label{solutiond4diag}
\ee
The sum extends over the different combinations of transverse 
momenta into two groups as they appear in (\ref{d4r}). 
In the picture of $t$-channel evolution the 
amplitudes start with the coupling of two reggeized gluons to the photons 
via a quark loop, then we have the propagation of the two--gluon 
state in the $t$-channel, and finally one of the gluons (or both) 
split --- or 'decay' --- 
into two or more gluons. To the splitting of the reggeized gluons 
belongs a certain color tensor, as given in (\ref{d4r}) for $D_4^R$. 
Viewed from a different angle, we can say that a group of gluons in the 
reggeizing part $D_4^R$ merges --- or 'collapses' --- 
to make up a more composite gluon 
which then enters the two--gluon compound state from below. 
We will use both pictures in parallel here and, depending on the context, 
speak of 'merging' or 'splitting' to mean the very same phenomenon 
of reggeization. 

We will now turn to the case of arbitrary $n$ and consider 
the reggeizing parts $D_n^R$ of the amplitudes. 
From these we derive the color 
structure accompanying the merging of a number of reggeized gluons 
into a single reggeon. 
This will lead us in a natural way to a simple classification of 
the composite reggeons according to their 
decay properties. Thereby we hope to gain a better understanding of  
how reggeization works. 
However, we have to keep in mind that the amplitudes $D_n^R$ 
constitute only the simplest part of our amplitudes and are 
derived from the special structure of the quark loop. 
In a later step we will eventually have 
to find out whether the reggeization 
in more complex amplitudes like $D_4^I$ works in the same way, 
that is whether it is accompanied by the same color tensors as in 
the two--gluon amplitude. 
Only then we may speak of a general property of the mechanism 
of reggeization. 

Let us pick one of the terms in the reggeizing part 
$D_n^R (\kf_1,\dots, \kf_n)$. 
It consists of a two--gluon (BFKL) amplitude $D_2$ with 
its two arguments made up from a group of 
the $n$ momenta each. 
Let us assume that the first of these groups contains 
$l$ gluons ($1\le l \le n-1$) and that the other group 
is made of the remaining $m$ gluons ($m=n-l$). 
For simplicity, we will further assume that the $l$ gluons 
in the first group are the first $l$ gluons in the amplitude 
with momenta $\kf_1, \dots, \kf_l$. 
Other terms in $D_n^R$ with a splitting into $l$ and $m$ 
gluons are then obtained by permutation of color and 
momentum labels. 
We can thus characterize the chosen term by its 
momentum structure, 
\be
 D_2(\,
\underbrace{\kf_1+\dots +\kf_l}_l \,,\, 
\underbrace{\kf_{l+1}+ \dots +\kf_n}_m \,)
\,.
\label{pickedtermmomentum}
\ee
In this section we will not care about the sign of the 
special term in $D_n^R$ we consider\footnote{Especially 
in the case of an odd number of gluons the relative 
signs of the terms in $D_n^R$ have to be treated with 
care since the signs change when the order of the color labels in 
the tensor is reversed.}. We will also neglect the 
additional factor $g^{n-2}$ that comes with the term above. 

To find the color tensor corresponding to (\ref{pickedtermmomentum}) 
for arbitrary $n$ we 
have to remind ourselves of the way the reggeizing parts $D_n^R$ 
were constructed. The individual terms in $D_n^R$ were 
obtained by the replacement $D_{(2;0)} \rarr D_2$ in 
the quark loop amplitude. The color tensor belonging to 
(\ref{pickedtermmomentum}) can therefore be deduced 
from the corresponding lowest order term in which $n$ gluons 
are coupled to the quark loop. Specifically, in the term of 
interest in the quark loop there are two contributions: 
one with the $l$ gluons of the first group coupled to the quark and 
the other $m$ gluons to the antiquark, the second with 
quark and antiquark exchanged (as described in section 
\ref{reviewmultigluon}). The trace in color 
space taken along the quark loop then gives for the first 
contribution 
\be
  \mbox{tr}(t^{b_1} \dots t^{b_l} t^{d_m} \dots t^{d_1}) 
\,.
\label{firstcontribcoltrace}
\ee
Here we have given new color labels to the gluons according 
to the group they are in. The first $l$ gluons now carry 
color labels $b_i$, the $m$ gluons in the second group 
have now been assigned the color labels $d_j$ such that the 
connection with the original labels is 
\be
 b_i = a_i \,\,\mbox{for}\,\, i \in \{ 1, \dots, l\} \,; 
\hspace{1cm}
 d_j = a_{l+j} \,\,\mbox{for}\,\, j \in \{ 1, \dots, m\} 
\,.
\ee
The second contribution contains a trace in color space in which 
the generators occur in reversed order, 
\be
  \mbox{tr}(t^{d_1} \dots t^{d_l} t^{b_m} \dots t^{b_1}) 
\,.
\label{secondcontribcoltrace}
\ee
The relative sign between the two contributions depends on the 
total number $n$ of gluons. (This is because the coupling 
of a gluon to a quark or antiquark in the quark loop effectively 
differ by a sign.) 
If $n$ is even, they come with the 
same sign. So the color tensor we are looking for is 
\be
 d^{b_1\dots b_ld_m\dots d_1}
\label{dtensorinterm}
\ee
as defined in (\ref{dallgdef}). If $n$ is odd, the two color traces 
in (\ref{firstcontribcoltrace}) and (\ref{secondcontribcoltrace}) 
come with opposite sign and we get a tensor of the form 
\be
 f^{b_1\dots b_ld_m\dots d_1}
\label{ftensorinterm}
\ee
as it was defined in (\ref{fallgdef}). 
It should be noted that in both cases the color labels of the 
one group come in ascending order in the tensor whereas those of the 
other group have to be taken in reversed order. 

Now we make a little digression. 
It will be useful to have a look at 
the color structure arising from the successive 
emission of $l$ gluons off a quark. 
In color space this process is associated with 
\be
 t^{b_1} \dots t^{b_l} = 
  2 \,\mbox{tr}(t^{b_1}\dots t^{b_l} t^c) t^c
 + \frac{1}{N_c} \,\mbox{tr}(t^{b_1}\dots t^{b_l} ) 
\,.
\label{loffquarkform}
\ee
The proof becomes almost obvious when we write this 
identity in birdtrack notation, 
\be
\label{loffquarkformdiag}
 \begin{picture}(0,0)%
\includegraphics{lanquark1.pstex}%
\end{picture}%
\setlength{\unitlength}{3947sp}%
\begingroup\makeatletter\ifx\SetFigFont\undefined
% extract first six characters in \fmtname
\def\x#1#2#3#4#5#6#7\relax{\def\x{#1#2#3#4#5#6}}%
\expandafter\x\fmtname xxxxxx\relax \def\y{splain}%
\ifx\x\y   % LaTeX or SliTeX?
\gdef\SetFigFont#1#2#3{%
  \ifnum #1<17\tiny\else \ifnum #1<20\small\else
  \ifnum #1<24\normalsize\else \ifnum #1<29\large\else
  \ifnum #1<34\Large\else \ifnum #1<41\LARGE\else
     \huge\fi\fi\fi\fi\fi\fi
  \csname #3\endcsname}%
\else
\gdef\SetFigFont#1#2#3{\begingroup
  \count@#1\relax \ifnum 25<\count@\count@25\fi
  \def\x{\endgroup\@setsize\SetFigFont{#2pt}}%
  \expandafter\x
    \csname \romannumeral\the\count@ pt\expandafter\endcsname
    \csname @\romannumeral\the\count@ pt\endcsname
  \csname #3\endcsname}%
\fi
\fi\endgroup
\begin{picture}(1019,622)(504,-286)
\put(976,-136){\makebox(0,0)[lb]{\smash{\SetFigFont{10}{12.0}{rm}$\dots$}}}
\put(1276,-286){\makebox(0,0)[lb]{\smash{\SetFigFont{10}{12.0}{rm}$b_l$}}}
\put(601,-286){\makebox(0,0)[lb]{\smash{\SetFigFont{10}{12.0}{rm}$b_1$}}}
\end{picture}
 = 
\,2 \,\begin{picture}(0,0)%
\includegraphics{lanquark2.pstex}%
\end{picture}%
\setlength{\unitlength}{3947sp}%
\begingroup\makeatletter\ifx\SetFigFont\undefined
% extract first six characters in \fmtname
\def\x#1#2#3#4#5#6#7\relax{\def\x{#1#2#3#4#5#6}}%
\expandafter\x\fmtname xxxxxx\relax \def\y{splain}%
\ifx\x\y   % LaTeX or SliTeX?
\gdef\SetFigFont#1#2#3{%
  \ifnum #1<17\tiny\else \ifnum #1<20\small\else
  \ifnum #1<24\normalsize\else \ifnum #1<29\large\else
  \ifnum #1<34\Large\else \ifnum #1<41\LARGE\else
     \huge\fi\fi\fi\fi\fi\fi
  \csname #3\endcsname}%
\else
\gdef\SetFigFont#1#2#3{\begingroup
  \count@#1\relax \ifnum 25<\count@\count@25\fi
  \def\x{\endgroup\@setsize\SetFigFont{#2pt}}%
  \expandafter\x
    \csname \romannumeral\the\count@ pt\expandafter\endcsname
    \csname @\romannumeral\the\count@ pt\endcsname
  \csname #3\endcsname}%
\fi
\fi\endgroup
\begin{picture}(719,790)(654,-454)
\put(1021,-346){\makebox(0,0)[lb]{\smash{\SetFigFont{10}{12.0}{rm}$\dots$}}}
\put(1332,-451){\makebox(0,0)[lb]{\smash{\SetFigFont{10}{12.0}{rm}$b_l$}}}
\put(654,-454){\makebox(0,0)[lb]{\smash{\SetFigFont{10}{12.0}{rm}$b_1$}}}
\end{picture}
 
+ \frac{1}{N_c} \,\begin{picture}(0,0)%
\includegraphics{lanquark3.pstex}%
\end{picture}%
\setlength{\unitlength}{3947sp}%
\begingroup\makeatletter\ifx\SetFigFont\undefined
% extract first six characters in \fmtname
\def\x#1#2#3#4#5#6#7\relax{\def\x{#1#2#3#4#5#6}}%
\expandafter\x\fmtname xxxxxx\relax \def\y{splain}%
\ifx\x\y   % LaTeX or SliTeX?
\gdef\SetFigFont#1#2#3{%
  \ifnum #1<17\tiny\else \ifnum #1<20\small\else
  \ifnum #1<24\normalsize\else \ifnum #1<29\large\else
  \ifnum #1<34\Large\else \ifnum #1<41\LARGE\else
     \huge\fi\fi\fi\fi\fi\fi
  \csname #3\endcsname}%
\else
\gdef\SetFigFont#1#2#3{\begingroup
  \count@#1\relax \ifnum 25<\count@\count@25\fi
  \def\x{\endgroup\@setsize\SetFigFont{#2pt}}%
  \expandafter\x
    \csname \romannumeral\the\count@ pt\expandafter\endcsname
    \csname @\romannumeral\the\count@ pt\endcsname
  \csname #3\endcsname}%
\fi
\fi\endgroup
\begin{picture}(719,790)(654,-454)
\put(1021,-346){\makebox(0,0)[lb]{\smash{\SetFigFont{10}{12.0}{rm}$\dots$}}}
\put(1332,-451){\makebox(0,0)[lb]{\smash{\SetFigFont{10}{12.0}{rm}$b_l$}}}
\put(654,-454){\makebox(0,0)[lb]{\smash{\SetFigFont{10}{12.0}{rm}$b_1$}}}
\end{picture}
 
\,.
\ee
To show this we recall that the decomposition
of a quark--antiquark state into a singlet and an adjoint
representation (known as the Fierz identity) is
\be
   \delta^\alpha_\gamma \delta^\delta_\beta
   = 2 (t^a)^\alpha_\beta (t^a)^\delta_\gamma
       + \fr {1}{N_c} \delta^\alpha_\beta \delta^\delta_\gamma\,,
\label{singoctlettintegr}
\ee
where $\alpha,\dots,\delta$ are color labels in the quark
(fundamental) representation.
In birdtracks it becomes
\be
   \begin{picture}(0,0)%
\includegraphics{singoct1.pstex}%
\end{picture}%
\setlength{\unitlength}{3947sp}%
\begingroup\makeatletter\ifx\SetFigFont\undefined
% extract first six characters in \fmtname
\def\x#1#2#3#4#5#6#7\relax{\def\x{#1#2#3#4#5#6}}%
\expandafter\x\fmtname xxxxxx\relax \def\y{splain}%
\ifx\x\y   % LaTeX or SliTeX?
\gdef\SetFigFont#1#2#3{%
  \ifnum #1<17\tiny\else \ifnum #1<20\small\else
  \ifnum #1<24\normalsize\else \ifnum #1<29\large\else
  \ifnum #1<34\Large\else \ifnum #1<41\LARGE\else
     \huge\fi\fi\fi\fi\fi\fi
  \csname #3\endcsname}%
\else
\gdef\SetFigFont#1#2#3{\begingroup
  \count@#1\relax \ifnum 25<\count@\count@25\fi
  \def\x{\endgroup\@setsize\SetFigFont{#2pt}}%
  \expandafter\x
    \csname \romannumeral\the\count@ pt\expandafter\endcsname
    \csname @\romannumeral\the\count@ pt\endcsname
  \csname #3\endcsname}%
\fi
\fi\endgroup
\begin{picture}(494,421)(804,-158)
\end{picture}
 = 2 \, \begin{picture}(0,0)%
\includegraphics{singoctoct.pstex}%
\end{picture}%
\setlength{\unitlength}{3947sp}%
\begingroup\makeatletter\ifx\SetFigFont\undefined
% extract first six characters in \fmtname
\def\x#1#2#3#4#5#6#7\relax{\def\x{#1#2#3#4#5#6}}%
\expandafter\x\fmtname xxxxxx\relax \def\y{splain}%
\ifx\x\y   % LaTeX or SliTeX?
\gdef\SetFigFont#1#2#3{%
  \ifnum #1<17\tiny\else \ifnum #1<20\small\else
  \ifnum #1<24\normalsize\else \ifnum #1<29\large\else
  \ifnum #1<34\Large\else \ifnum #1<41\LARGE\else
     \huge\fi\fi\fi\fi\fi\fi
  \csname #3\endcsname}%
\else
\gdef\SetFigFont#1#2#3{\begingroup
  \count@#1\relax \ifnum 25<\count@\count@25\fi
  \def\x{\endgroup\@setsize\SetFigFont{#2pt}}%
  \expandafter\x
    \csname \romannumeral\the\count@ pt\expandafter\endcsname
    \csname @\romannumeral\the\count@ pt\endcsname
  \csname #3\endcsname}%
\fi
\fi\endgroup
\begin{picture}(480,255)(1486,-76)
\end{picture}

    + \fr{1}{N_c}\, \begin{picture}(0,0)%
\includegraphics{singoctsing.pstex}%
\end{picture}%
\setlength{\unitlength}{3947sp}%
\begingroup\makeatletter\ifx\SetFigFont\undefined
% extract first six characters in \fmtname
\def\x#1#2#3#4#5#6#7\relax{\def\x{#1#2#3#4#5#6}}%
\expandafter\x\fmtname xxxxxx\relax \def\y{splain}%
\ifx\x\y   % LaTeX or SliTeX?
\gdef\SetFigFont#1#2#3{%
  \ifnum #1<17\tiny\else \ifnum #1<20\small\else
  \ifnum #1<24\normalsize\else \ifnum #1<29\large\else
  \ifnum #1<34\Large\else \ifnum #1<41\LARGE\else
     \huge\fi\fi\fi\fi\fi\fi
  \csname #3\endcsname}%
\else
\gdef\SetFigFont#1#2#3{\begingroup
  \count@#1\relax \ifnum 25<\count@\count@25\fi
  \def\x{\endgroup\@setsize\SetFigFont{#2pt}}%
  \expandafter\x
    \csname \romannumeral\the\count@ pt\expandafter\endcsname
    \csname @\romannumeral\the\count@ pt\endcsname
  \csname #3\endcsname}%
\fi
\fi\endgroup
\begin{picture}(480,255)(1486,-76)
\end{picture}

\,.
\label{singoctbirdintegr}
\ee 
Applying this identity to the 
right hand side of (\ref{loffquarkformdiag}) we immediately 
find the left hand side. 
Using further the definitions (\ref{dallgdef}), (\ref{fallgdef})
of $d$ and $f$ tensors we can rewrite this as 
\be
t^{b_1} \dots t^{b_l} = 
\left[ d^{b_1 \dots b_l c} + i f^{b_1 \dots b_l c}
\right] t^c
+ \frac{1}{2 N_c} 
\left[ d^{b_1 \dots b_l } + i f^{b_1 \dots b_l }
\right] 
\,.
\label{loffquarkformdf}
\ee
A special case of that general fact 
is the well--known formula describing the 
successive emission of two gluons off a quark, 
\be
t^a t^b = \frac{1}{2} \left[ \frac{1}{N_c} \delta_{ab} 
 + (d_{abc} + i f_{abc}) t^c \right] 
\,.
\label{GMab}
\ee
Here we have to have in mind that the conventional normalization 
of structure constants $f_{abc}$ and $d_{abc}$ differs from 
our definition of $d$- and $f$-tensors with upper indices. 
Further, we have to use $d^{ab}= \delta_{ab}$ and 
$f^{ab}=0$. 
A seemingly trivial special case 
of (\ref{loffquarkformdf}) is the emission of a single gluon. 
Namely, if $l=1$ only the first term on the right hand side 
gives a contribution, and due to $d^{ab}= \delta_{ab}$ we 
end up with a trivial identity. Nevertheless, this case is quite 
important for the consistency of the assignment of 'quantum 
numbers' we want to carry out. 

Coming back to our main problem now, we apply these 
identities to the color tensor of the term we have picked in $D_n^R$. 
The color tensor of that term depends on the total number $n$ 
of gluons. For even $n$ the color tensor 
associated with our term is the $d$-tensor 
in (\ref{dtensorinterm}). Applying (\ref{loffquarkformdf}) 
now to the first group containing $l$ gluons we arrive at 
\bea
\label{reggetensorsausd}
d^{b_1\dots b_ld_m\dots d_1} &=& 
d^{b_1 \dots b_l c} d^{cd_1 \dots d_m} 
+ f^{b_1 \dots b_l c} f^{cd_1 \dots d_m} \nn \\
&& +\frac{1}{2 N_c} \,d^{b_1 \dots b_l} d^{d_1 \dots d_m}
+\frac{1}{2 N_c} f^{b_1 \dots b_l} f^{d_1 \dots d_m}
\,.
\eea
For odd $n$ the color tensor is the $f$-tensor 
in (\ref{ftensorinterm}) and here the application 
of (\ref{loffquarkformdf}) to the $l$ gluons in the 
first group gives 
\bea
\label{reggetensorsausf}
f^{b_1\dots b_ld_m\dots d_1} &=& 
- \,d^{b_1 \dots b_l c} f^{cd_1 \dots d_m} 
+ f^{b_1 \dots b_l c} d^{cd_1 \dots d_m} \nn \\
&& -\frac{1}{2 N_c} \,d^{b_1 \dots b_l} f^{d_1 \dots d_m}
+\frac{1}{2 N_c} f^{b_1 \dots b_l} d^{d_1 \dots d_m}
\,.
\eea
The relative signs again depend on the order 
of gluons we started with due to the definition of the $f$-tensor. 
We will not pay special attention to this detail here and 
concentrate on the tensors in the four terms separately. 

These two decompositions of the $d$- and $f$-tensors 
contain all possible combinations of even and odd 
$l$ and $m$, and the 
reggeization tensors we are looking for can now 
be extracted from the two decompositions. 
Obviously, the tensors describing the splitting in the 
two groups are correlated. We will now try to assign 
a kind of quantum number to the two reggeized gluons 
according to the way the respective reggeons split. 
This can only make sense if we demand that the two 
gluons in the two--reggeon state 
carry the same 'quantum number'. 
There will be four possible types of reggeons. We first want 
to fix the type of a reggeized gluon with color label $c$ 
that does not split. 
This case is included in the above identities as the 
splitting into one gluon ($l=1$ or $m=1$) via a 
$d$-tensor (since we had $d^{ab}=\delta_{ab}$). 
We want to call this type a reggeized gluon of type $f$ in 
the adjoint representation. Now we can read off from 
the first term in (\ref{reggetensorsausd}) that such a 
reggeon decays 
into an odd number of gluons with a $d$-tensor. From 
the first or second term in (\ref{reggetensorsausf}) 
we find that it decays into an even number of gluons 
with an $f$-tensor. This assignment is consistent 
also if both $l>1$ and $m>1$. 

Secondly, there is also the possibility that a reggeized gluon 
with color label $c$ 
splits into an even number of gluons via a $d$-tensor. 
We want to call such a reggeized gluon a reggeon 
of type $d$ in 
the adjoint representation. From the first lines in 
(\ref{reggetensorsausd}) and (\ref{reggetensorsausf}) 
we find consistently that such a reggeon splits into 
an odd number of gluons via an $f$-tensor. 
An interesting observation is that a reggeon of this type 
can only occur if it is composite of at least two gluons. 
Otherwise the corresponding terms in the above 
identities vanish due to $f^{ab} = 0$. 
Consequently, such a reggeon can be found in the two--gluon 
state only if both reggeized gluons decay. 

Now we proceed to the last two terms in the right hand sides 
of (\ref{reggetensorsausd}) and (\ref{reggetensorsausf}). 
Here we observe that each of the two groups of reggeons 
has zero total color charge. The corresponding composite 
reggeon is obviously in a color singlet state. 
Although this appears to be somewhat counter--intuitive 
on first sight we will try to treat these singlet reggeons 
on equal footing with the reggeons in the adjoint 
representation discussed before. 
Again, we can consistently define two different types, 
$d$ and $f$. 
A singlet reggeon of type $f$ splits into an even 
number of gluons via an $f$-tensor and into an odd number 
of gluons via a $d$-tensor. 
A $d$-type singlet reggeon decays into an even number of 
gluons via a $d$-tensor and into an odd number of 
gluons via an $f$-tensor. 
Here we do not have a certain decay mode we 
want to fix as of type $d$ or $f$ like in the case of the 
adjoint representation. Therefore in the assignment 
we could as well interchange $d$ and $f$. 
Like in the case of the $d$-type reggeon in the 
adjoint representation the singlet reggeons are composite 
of at least two gluons. They cannot occur in a term in $D_n^R$ 
in which only one of the two reggeized gluons in the 
two--gluon compound state decays. 

We have been able to extract from the two color identities 
(\ref{reggetensorsausd}) and (\ref{reggetensorsausf}) 
in a consistent way a classification of reggeized gluons 
in the reggeizing amplitudes $D_n^R$. 
The classification is valid for arbitrary $n$ and all 
possible combinations of numbers $l$ and $m$ that 
merge into one of the two reggeized gluons entering 
the two--gluon state. 
Let us summarize the assignments of reggeon types 
and the corresponding reggeization tensors in 
table \ref{reggetable}. 
Here $l$ denotes the number of reggeized gluons that merge into 
a more composite one. 
The diagrams for the color tensors are drawn with 
four and three legs here for illustration only. 
%\begin{table}[htbp]
%\begin{center}
\TABLE{
\renewcommand{\arraystretch}{2}
\setlength{\tabcolsep}{.7cm}
\begin{tabular}{|c||c|c|} \hline
\rule[-4mm]{0cm}{1cm} 
reggeon type  & \rule{5cm}{0cm} \hspace{-5cm}
$l$ even & $l$ odd \\ \hline 
\rule[-6mm]{0cm}{1.4cm}
$f$, adjoint rep.\ &   \begin{picture}(0,0)%
\includegraphics{f1234c.pstex}%
\end{picture}%
\setlength{\unitlength}{3947sp}%
\begingroup\makeatletter\ifx\SetFigFont\undefined
% extract first six characters in \fmtname
\def\x#1#2#3#4#5#6#7\relax{\def\x{#1#2#3#4#5#6}}%
\expandafter\x\fmtname xxxxxx\relax \def\y{splain}%
\ifx\x\y   % LaTeX or SliTeX?
\gdef\SetFigFont#1#2#3{%
  \ifnum #1<17\tiny\else \ifnum #1<20\small\else
  \ifnum #1<24\normalsize\else \ifnum #1<29\large\else
  \ifnum #1<34\Large\else \ifnum #1<41\LARGE\else
     \huge\fi\fi\fi\fi\fi\fi
  \csname #3\endcsname}%
\else
\gdef\SetFigFont#1#2#3{\begingroup
  \count@#1\relax \ifnum 25<\count@\count@25\fi
  \def\x{\endgroup\@setsize\SetFigFont{#2pt}}%
  \expandafter\x
    \csname \romannumeral\the\count@ pt\expandafter\endcsname
    \csname @\romannumeral\the\count@ pt\endcsname
  \csname #3\endcsname}%
\fi
\fi\endgroup
\begin{picture}(545,474)(930,-748)
\put(1160,-546){\makebox(0,0)[lb]{\smash{\SetFigFont{8}{9.6}{rm}$f$}}}
\end{picture}
&   
\begin{picture}(0,0)%
\includegraphics{odddc.pstex}%
\end{picture}%
\setlength{\unitlength}{3947sp}%
\begingroup\makeatletter\ifx\SetFigFont\undefined
% extract first six characters in \fmtname
\def\x#1#2#3#4#5#6#7\relax{\def\x{#1#2#3#4#5#6}}%
\expandafter\x\fmtname xxxxxx\relax \def\y{splain}%
\ifx\x\y   % LaTeX or SliTeX?
\gdef\SetFigFont#1#2#3{%
  \ifnum #1<17\tiny\else \ifnum #1<20\small\else
  \ifnum #1<24\normalsize\else \ifnum #1<29\large\else
  \ifnum #1<34\Large\else \ifnum #1<41\LARGE\else
     \huge\fi\fi\fi\fi\fi\fi
  \csname #3\endcsname}%
\else
\gdef\SetFigFont#1#2#3{\begingroup
  \count@#1\relax \ifnum 25<\count@\count@25\fi
  \def\x{\endgroup\@setsize\SetFigFont{#2pt}}%
  \expandafter\x
    \csname \romannumeral\the\count@ pt\expandafter\endcsname
    \csname @\romannumeral\the\count@ pt\endcsname
  \csname #3\endcsname}%
\fi
\fi\endgroup
\begin{picture}(324,474)(1039,-748)
\put(1160,-553){\makebox(0,0)[lb]{\smash{\SetFigFont{8}{9.6}{rm}$d$}}}
\end{picture}
   \\ 
 \rule[-6mm]{0cm}{1.4cm}
$d$, adjoint rep.\ &   \begin{picture}(0,0)%
\includegraphics{d1234c.pstex}%
\end{picture}%
\setlength{\unitlength}{3947sp}%
\begingroup\makeatletter\ifx\SetFigFont\undefined
% extract first six characters in \fmtname
\def\x#1#2#3#4#5#6#7\relax{\def\x{#1#2#3#4#5#6}}%
\expandafter\x\fmtname xxxxxx\relax \def\y{splain}%
\ifx\x\y   % LaTeX or SliTeX?
\gdef\SetFigFont#1#2#3{%
  \ifnum #1<17\tiny\else \ifnum #1<20\small\else
  \ifnum #1<24\normalsize\else \ifnum #1<29\large\else
  \ifnum #1<34\Large\else \ifnum #1<41\LARGE\else
     \huge\fi\fi\fi\fi\fi\fi
  \csname #3\endcsname}%
\else
\gdef\SetFigFont#1#2#3{\begingroup
  \count@#1\relax \ifnum 25<\count@\count@25\fi
  \def\x{\endgroup\@setsize\SetFigFont{#2pt}}%
  \expandafter\x
    \csname \romannumeral\the\count@ pt\expandafter\endcsname
    \csname @\romannumeral\the\count@ pt\endcsname
  \csname #3\endcsname}%
\fi
\fi\endgroup
\begin{picture}(545,474)(930,-748)
\put(1160,-553){\makebox(0,0)[lb]{\smash{\SetFigFont{8}{9.6}{rm}$d$}}}
\end{picture}
& 
\begin{picture}(0,0)%
\includegraphics{oddfc.pstex}%
\end{picture}%
\setlength{\unitlength}{3947sp}%
\begingroup\makeatletter\ifx\SetFigFont\undefined
% extract first six characters in \fmtname
\def\x#1#2#3#4#5#6#7\relax{\def\x{#1#2#3#4#5#6}}%
\expandafter\x\fmtname xxxxxx\relax \def\y{splain}%
\ifx\x\y   % LaTeX or SliTeX?
\gdef\SetFigFont#1#2#3{%
  \ifnum #1<17\tiny\else \ifnum #1<20\small\else
  \ifnum #1<24\normalsize\else \ifnum #1<29\large\else
  \ifnum #1<34\Large\else \ifnum #1<41\LARGE\else
     \huge\fi\fi\fi\fi\fi\fi
  \csname #3\endcsname}%
\else
\gdef\SetFigFont#1#2#3{\begingroup
  \count@#1\relax \ifnum 25<\count@\count@25\fi
  \def\x{\endgroup\@setsize\SetFigFont{#2pt}}%
  \expandafter\x
    \csname \romannumeral\the\count@ pt\expandafter\endcsname
    \csname @\romannumeral\the\count@ pt\endcsname
  \csname #3\endcsname}%
\fi
\fi\endgroup
\begin{picture}(324,474)(1039,-748)
\put(1160,-538){\makebox(0,0)[lb]{\smash{\SetFigFont{8}{9.6}{rm}$f$}}}
\end{picture}
 \\ 
 \rule[-6mm]{0cm}{1.4cm}
$f$, singlet &   \begin{picture}(0,0)%
\includegraphics{f1234.pstex}%
\end{picture}%
\setlength{\unitlength}{3947sp}%
\begingroup\makeatletter\ifx\SetFigFont\undefined
% extract first six characters in \fmtname
\def\x#1#2#3#4#5#6#7\relax{\def\x{#1#2#3#4#5#6}}%
\expandafter\x\fmtname xxxxxx\relax \def\y{splain}%
\ifx\x\y   % LaTeX or SliTeX?
\gdef\SetFigFont#1#2#3{%
  \ifnum #1<17\tiny\else \ifnum #1<20\small\else
  \ifnum #1<24\normalsize\else \ifnum #1<29\large\else
  \ifnum #1<34\Large\else \ifnum #1<41\LARGE\else
     \huge\fi\fi\fi\fi\fi\fi
  \csname #3\endcsname}%
\else
\gdef\SetFigFont#1#2#3{\begingroup
  \count@#1\relax \ifnum 25<\count@\count@25\fi
  \def\x{\endgroup\@setsize\SetFigFont{#2pt}}%
  \expandafter\x
    \csname \romannumeral\the\count@ pt\expandafter\endcsname
    \csname @\romannumeral\the\count@ pt\endcsname
  \csname #3\endcsname}%
\fi
\fi\endgroup
\begin{picture}(545,342)(930,-748)
\put(1160,-542){\makebox(0,0)[lb]{\smash{\SetFigFont{8}{9.6}{rm}$f$}}}
\end{picture}
 &    
\begin{picture}(0,0)%
\includegraphics{oddd.pstex}%
\end{picture}%
\setlength{\unitlength}{3947sp}%
\begingroup\makeatletter\ifx\SetFigFont\undefined
% extract first six characters in \fmtname
\def\x#1#2#3#4#5#6#7\relax{\def\x{#1#2#3#4#5#6}}%
\expandafter\x\fmtname xxxxxx\relax \def\y{splain}%
\ifx\x\y   % LaTeX or SliTeX?
\gdef\SetFigFont#1#2#3{%
  \ifnum #1<17\tiny\else \ifnum #1<20\small\else
  \ifnum #1<24\normalsize\else \ifnum #1<29\large\else
  \ifnum #1<34\Large\else \ifnum #1<41\LARGE\else
     \huge\fi\fi\fi\fi\fi\fi
  \csname #3\endcsname}%
\else
\gdef\SetFigFont#1#2#3{\begingroup
  \count@#1\relax \ifnum 25<\count@\count@25\fi
  \def\x{\endgroup\@setsize\SetFigFont{#2pt}}%
  \expandafter\x
    \csname \romannumeral\the\count@ pt\expandafter\endcsname
    \csname @\romannumeral\the\count@ pt\endcsname
  \csname #3\endcsname}%
\fi
\fi\endgroup
\begin{picture}(324,337)(1039,-748)
\put(1160,-553){\makebox(0,0)[lb]{\smash{\SetFigFont{8}{9.6}{rm}$d$}}}
\end{picture}
   \\ 
 \rule[-6mm]{0cm}{1.4cm}
$d$, singlet &   \begin{picture}(0,0)%
\includegraphics{d1234neu.pstex}%
\end{picture}%
\setlength{\unitlength}{3947sp}%
\begingroup\makeatletter\ifx\SetFigFont\undefined
% extract first six characters in \fmtname
\def\x#1#2#3#4#5#6#7\relax{\def\x{#1#2#3#4#5#6}}%
\expandafter\x\fmtname xxxxxx\relax \def\y{splain}%
\ifx\x\y   % LaTeX or SliTeX?
\gdef\SetFigFont#1#2#3{%
  \ifnum #1<17\tiny\else \ifnum #1<20\small\else
  \ifnum #1<24\normalsize\else \ifnum #1<29\large\else
  \ifnum #1<34\Large\else \ifnum #1<41\LARGE\else
     \huge\fi\fi\fi\fi\fi\fi
  \csname #3\endcsname}%
\else
\gdef\SetFigFont#1#2#3{\begingroup
  \count@#1\relax \ifnum 25<\count@\count@25\fi
  \def\x{\endgroup\@setsize\SetFigFont{#2pt}}%
  \expandafter\x
    \csname \romannumeral\the\count@ pt\expandafter\endcsname
    \csname @\romannumeral\the\count@ pt\endcsname
  \csname #3\endcsname}%
\fi
\fi\endgroup
\begin{picture}(545,342)(930,-748)
\put(1160,-553){\makebox(0,0)[lb]{\smash{\SetFigFont{8}{9.6}{rm}$d$}}}
\end{picture}
 & 
\begin{picture}(0,0)%
\includegraphics{oddf.pstex}%
\end{picture}%
\setlength{\unitlength}{3947sp}%
\begingroup\makeatletter\ifx\SetFigFont\undefined
% extract first six characters in \fmtname
\def\x#1#2#3#4#5#6#7\relax{\def\x{#1#2#3#4#5#6}}%
\expandafter\x\fmtname xxxxxx\relax \def\y{splain}%
\ifx\x\y   % LaTeX or SliTeX?
\gdef\SetFigFont#1#2#3{%
  \ifnum #1<17\tiny\else \ifnum #1<20\small\else
  \ifnum #1<24\normalsize\else \ifnum #1<29\large\else
  \ifnum #1<34\Large\else \ifnum #1<41\LARGE\else
     \huge\fi\fi\fi\fi\fi\fi
  \csname #3\endcsname}%
\else
\gdef\SetFigFont#1#2#3{\begingroup
  \count@#1\relax \ifnum 25<\count@\count@25\fi
  \def\x{\endgroup\@setsize\SetFigFont{#2pt}}%
  \expandafter\x
    \csname \romannumeral\the\count@ pt\expandafter\endcsname
    \csname @\romannumeral\the\count@ pt\endcsname
  \csname #3\endcsname}%
\fi
\fi\endgroup
\begin{picture}(324,337)(1039,-748)
\put(1160,-542){\makebox(0,0)[lb]{\smash{\SetFigFont{8}{9.6}{rm}$f$}}}
\end{picture}
 \\ \hline
\end{tabular} 
\renewcommand{\arraystretch}{1}
\caption{Reggeization tensors as obtained from the 
reggeizing parts $D_n^R$}
\label{reggetable}
}
%\end{center}
%\end{table}
Of course they 
represent the tensors of type $d$ or $f$ for an arbitrary 
number $l$ of gluons as defined in (\ref{dallgdef}) 
and (\ref{fallgdef}). 
For $l \ge 3$ ($l \ge 4$ for the singlet reggeons) 
the $d$- and $f$-tensors in the table can 
be decomposed further into contractions of symmetric and 
antisymmetric structure constants. 
The corresponding formulae for up to $l=5$ ($l = 6$ for 
the singlet reggeons) can be found in \cite{BE}. 

In the case of $l=2$ the $d$-type reggeons 
can be interpreted as symmetric in the two gluons, and 
the $f$-type reggeons are antisymmetric in the two gluons. 
This holds in the adjoint representation as well as in the 
color singlet. However, this interpretation 
of $d$ ($f$) as symmetric (antisymmetric) 
has to be refined for $l\ge 3$ since then the tensors 
are not completely (anti)symmetric in the $l$ color 
labels. Instead the symmetry of the $d$- and $f$-tensors 
in individual pairs of gluons becomes more complicated. 

In the reggeizing parts $D_n^R$ of the $n$-gluon amplitudes 
the different types of reggeons in our classification occur 
inevitably at the same time, since the tensors  
(\ref{reggetensorsausd}) and (\ref{reggetensorsausf}) 
contain them together. This is not necessarily the case 
in more complicated parts of the amplitudes that contain 
a compound state of more than two reggeons. 
For example, higher amplitudes will contain a part that does not 
reggeize with respect to the two--gluon amplitude 
but does reggeize with respect to the four--gluon amplitude. 
An example is the part $D_5^I$ of the five--gluon 
amplitude. It is well conceivable that in such amplitudes 
the four reggeons in the compound state are less 
correlated than in the two--gluon state and the 
reggeon types can occur independently. 

The example of the part $D_5^I$ of the five--gluon amplitude 
is, however, not sufficient to clarify the situation. 
On the one hand, it confirms our classification: 
the four gluons in $D_4^I$ are of type $f$ in the 
adjoint representation and one of them in fact splits 
into two with the tensor that should be expected from 
our classification. 
On the other hand, we do not expect the other types of reggeons 
to appear in $D_5^I$ if our classification is right, since those 
require two splittings in the whole amplitude. 
We thus have to go at least to 
the six--gluon amplitude to study their behavior. 

In this respect 
the $f$-type reggeon in the adjoint representation 
plays a special role. It is the only type of reggeon 
that occurs in $D_n^R$ when only one of the two reggeons 
decays, i.\,e.\ when the other one splits trivially into one gluon. 
This observation leads us to suspect that the $f$-type 
reggeon in the adjoint representation can appear in each possible 
compound state. 
For the other types the correlation of the two reggeons in the 
two--gluon compound state seems to be essential. 
It seems natural to expect that their behavior in 
higher compound states is more complicated. 
Most probably the knowledge of the reggeizing parts $D_n^R$ 
is not sufficient to fully understand these types. 

At present, the classification developed in this section 
has the status of an observation. We are not able to 
derive the decay tensors from first principles for arbitrary reggeons 
in an arbitrary amplitude. 
Certainly, our findings have to be tested in 
the investigation of higher $n$-gluon amplitudes. 
Especially the universality of the concept outlined here 
is by no means obvious. 

We should note that there is probably a deeper connection 
with the notion of signature (see for example 
\cite{signature,Bartelsnuclphys}) 
to be discovered here. The reggeon of type $f$ in the adjoint 
representation can be identified as a reggeized gluon with 
the usual negative signature. 
The reggeon of type $d$ in the adjoint representation 
can probably be identified with a reggeized gluon 
carrying positive signature. 
For the singlet reggeons, however, the situation is less clear. 
Presumably it will be necessary to clarify their meaning 
before the full relation to signature can be established. 

\section{Conclusions}
\label{summary}

We have used the generalized leading logarithmic approximation (GLLA) 
in perturbative QCD to study the reggeization of the gluon 
at high energies. The reggeized gluon represents an infinite 
sum of Feynman diagrams in which gluons arrange 
in such a way that they form a collective excitation of the 
gauge field with gluon quantum numbers. 
This mechanism is reflected in the properties 
of the $n$-gluon amplitudes arising in the framework of the GLLA. 
These amplitudes describe the production of $n$ gluons in the 
$t$-channel. They are known explicitly for up to $n=5$ gluons, 
and the so--called reggeizing part of the amplitudes can be 
computed even for arbitrary $n$. Some parts of the $n$-gluon 
amplitudes turn out to be superpositions of lower amplitudes 
in which a group of gluons merges to form a composite object, 
i.\,e.\ a reggeized gluon. In this way the amplitudes contain 
information on how the formation of the reggeized gluon 
takes place. 

The reggeization of the gluon is closely related to the 
emergence of the field theory structure in the 
$n$-gluon amplitudes. In this structure there are 
$n$-gluon states in the $t$-channel with even $n$ only 
which are coupled to each other by transition vertices 
like the two--to--four or the two--to--six gluon vertex. 
Accordingly, the amplitudes consist of different parts 
which are superpositions of two--gluon states, 
irreducible four--gluon states etc. 

We have found Ward type identities relating $n$-gluon amplitudes 
with different numbers of gluons. They apply when one of the 
transverse momenta of the gluons vanishes. These identities 
place strong constraints on the complicated interplay of the 
color and momentum structure in the amplitudes which 
reflects the non--abelian character of QCD. 
The Ward type identities have been extracted from the 
known solutions of the $n$-gluon amplitudes. We have 
formulated conjectures about their generalization to higher $n$. 
An important point is that the Ward type identities apply 
to different parts of the amplitudes separately. 
For the reggeizing part (the part that is a superposition of 
two--gluon amplitudes) we have been able to prove the 
identities for arbitrary $n$. We expect them to 
hold separately also for those parts that are superpositions of 
irreducible four--gluon and six--gluon amplitudes etc., 
and we have shown that this is true in the five--gluon 
amplitude. 
They will therefore be useful for the difficult problem 
of separating these parts in higher amplitudes ($n\ge 6$), 
and consequently in identifying the elements of the 
effective field theory. 

The Ward type identities found here have some similarity 
to relations for certain families of $n$-reggeon states 
which have recently been discovered \cite{Vacca2}. 
Those relations hold for states with a fixed number of reggeized gluons 
in the large-$N_c$ limit which are 
described by the BKP equations \cite{Bartelskernels,BKP}. 
The relations in \cite{Vacca2} allow one to construct an 
$n+1$-reggeon state from a state with $n$ reggeons, thus 
relating states with different numbers of reggeized gluons. 
Although those relations and our Ward type identities apply 
to somewhat different amplitudes they might possibly have 
a common origin. 

The reggeizing part of each $n$-gluon amplitude is a 
superposition of two--gluon (BFKL) amplitudes in which a number 
of gluons merges to form a reggeized gluon. We have identified 
the tensors associated with this process in color space. 
The color tensors carry the main information about reggeization 
in the two--gluon amplitude. 
It was possible to find these tensors for arbitrary numbers 
of gluons forming a more composite reggeized gluon. 
It appears possible to assign a sort of quantum number to 
the reggeized gluon according to how it is formed in 
color space. The possible types of reggeized gluons include 
two reggeized gluons in the adjoint representation and two 
in a color singlet state. 
We have addressed the possible universality of these color 
tensors, i.\,e.\ the question whether they can also occur 
in more complicated amplitudes like the 
irreducible four--gluon state. We expect that one of the 
reggeon types (the $f$-reggeon in the adjoint representation) 
can occur in arbitrary amplitudes. This is confirmed by 
the explicit solution for the five--gluon amplitude. 
The behaviour of the other types might be more involved 
in higher amplitudes because their occurrence requires correlations between 
different reggeized gluons. It would obviously be very 
interesting to test the universality of this classification 
by studying the relevant parts of the six--gluon amplitude, 
namely the ones involving an irreducible four--gluon state. 

With our results we hope to have made a step towards 
a better understanding not only of reggeization itself 
but also of the effective field theory structure of unitarity 
corrections. 
There is a variety of other approaches to unitarity corrections 
in high energy QCD (for an extensive list of references see \cite{BE}). 
In many of them the reggeized gluon plays a prominent r\^ole. 
It would of course be desirable to confirm the conjectures made 
in the present paper using one of these approaches. 
A promising starting point might for example be 
the effective action approach of \cite{effaction,Lipatovphysrep}. 
A crucial test of 
the conjectures will also be the further investigation of the irreducible 
part of the six--gluon amplitude. 

\section*{Acknowledgments}
I would like to thank Jochen Bartels for interesting discussions.

\end{document}